\documentclass[10pt]{article}

\usepackage{times}
\usepackage{amsmath}
\usepackage{amsfonts}
\usepackage{bm}
\usepackage[numbers,square,sort&compress]{natbib}
\usepackage{graphicx}
\usepackage{subfigure}
\usepackage{enumerate}
\usepackage{color}
\usepackage{algorithm}
\usepackage{algorithmic}

\begin{document}

\title{Viscous regularization and r-adaptive remeshing for finite
  element analysis of lipid membrane mechanics}

\author{Lin Ma and 
William S. Klug\footnote{Corresponding author: klug@ucla.edu}\\
\normalsize Department of Mechanical and Aerospace Engineering, \\
\normalsize University of California, Los Angeles, CA, 90095, United States}

\date{\today}

\maketitle

\begin{abstract}
As two-dimensional fluid shells, lipid bilayer membranes resist
bending and stretching but are unable to sustain shear stresses.  This
property gives membranes the ability to adopt dramatic shape
changes. In this paper, a finite element model is developed to study
static equilibrium mechanics of membranes. In particular, a viscous
regularization method is proposed to stabilize tangential mesh
deformations and improve the convergence rate of nonlinear solvers.
The Augmented Lagrangian method is used to enforce global constraints
on area and volume during membrane deformations. As a validation of
the method, equilibrium shapes for a shape-phase diagram of lipid
bilayer vesicle are calculated.  These numerical techniques are also
shown to be useful for simulations of three-dimensional
large-deformation problems: the formation of tethers (long tube-like
exetensions); and Ginzburg-Landau phase separation of a
two-lipid-component vesicle.  To deal with the large mesh distortions
of the two-phase model, modification of vicous regularization is
explored to achieve r-adaptive mesh optimization.
\end{abstract}

\section{Introduction}\label{sec:intro}
Lipid membranes are a critical part of life because they serve as a
barrier to separate the contents of the cell from the external
world. Lipid molecules are composed of a hydrophilic headgroup and two
hydrophobic hydrocarbon chains \cite{AlbertsECB}, and will form a
bilayer structure spontaneously by the hydrophobic effect when
introduced into water in sufficient concentration.  Though the cell
membrane has more complex structure, being littered with all kinds of
proteins that serve as selective receptors, channels, and pumps, in
this paper we will focus on closed spherical pure lipid bilayer
membranes, i.e., vesicles.

Common experience reveals that it is much easier to bend a thin plate
than to stretch it (a good example is a sheet of paper). This is also
true for lipid bilayer membranes, except that there's no shear force
because of the fluid property of membranes. The mechanical energy of a
lipid bilayer has three major contributors: bending (curvature) of
each monolayer; area or in-plane expansion and contraction of each
monolayer; and osmotic pressure. Because the last two energy scales
are much larger than the first one (by several orders of magnitude)
\cite{Seifert1997}, they effectively place constraints on the total
surface area and enclosed volume of the bilayer membrane on
experimental time scales (up to at least one hour).  Thus, the
mechanically interesting energy arises from bending of the membrane.

Canham \cite{Canham1970}, Helfrich \cite{Helfrich1973} and Evans
\cite{Evans1974} pioneered the development of the lowest-order bending
energy theory, often referred to as the spontaneous curvature model, in
which energy is a quadratic function of the principle curvatures and
the intrinsic or \emph{spontaneous} curvature of the
surface. Incremental improvements to this model include the bilayer
couple model \cite{Svetina1982, Svetina1983}, which imposes the hard
constraint on the area difference of the two monolayers, and the
area-difference-elasticity (ADE) model \cite{Seifert1992, Wiese1992,
Bozic1992}, which adds a non-local curvature energy term representing
an elastic penalty on the area difference.

The equations of equilibrium for the spontaneous curvature model,
first calculated by Jenkins \cite{Jenkins1977a, Jenkins1977b}, are
difficult to solve being highly nonlinear fourth-order PDEs. 
The most common approach to modeling membrane mechanics numerically
has been to discretize a vesicle surface by a triangle mesh, and
approximate the curvature along mesh edges with finite-difference (FD)
operators. Starting from some suitable initial shape, the FD
approximation of curvature energy can be summed on the triangulation,
and an adjacent local minimum then can be found by downhill
minimization (often via a conjugate gradient algorithm) \cite{Hsu1992,
Jaric1995, Kraus1995, Wintz1996}.  Another mesh-based approach, the
finite element method (FEM), was also recently applied to the study of
membrane mechanics by Feng and Klug \cite{Feng2006}, using
$C^1$-conforming triangular subdivision-surfaces elements to
approximate the membrane curvature energy.  

One feature shared in common among these mesh-based methods is the
need for stabilization of mesh vertex motions tangential to the
discretized surface.  This issue arises as a fundamental consequence
of the use of a mesh for explicit coordinate parameterization of the
geometry of a fluid membrane having no physically meaningful reference
configuration.  As pointed out previously
\citep[e.g.,][]{GompperKroll1997, Capovilla2003}, the dependence of
the curvature energy functional on the surface position map is
invariant upon changes in parameterization.  Physically this implies
that in-plane dilatational and shear modes of mesh deformation carry
no energy cost, and therefore, no stiffness.  The addition of an
artificial in-plane stiffness, for example by placing Hookean springs
along the edges of a triangular mesh \citep{GompperKroll1997} does
indeed stabilize these motions, but in doing so also changes the
physics, yielding a model for a polymerized (rather than fluid)
membrane.  In order to allow for fluid-like diffusion of membrane
vertices, a number of researchers have used a \emph{dynamic
  triangulation} approach, wherein the edge connecting a pair of
adjacent triangles along one of the diagonals between the four
associated vertices is swapped for the other diagonal.  Within a Monte
Carlo simulation framework this method yields mean-square vertex
displacements that are consistent with microscopic diffusion
\citep{HoBaumgartner1990}, and this approach has been shown to produce
an effective viscosity that increases as the edge swapping rate
decreases \citep{NoguchiGompper2005}.  However, it is unclear whether
the dynamic triangulation approach can enable unphysical in-plane
forces to fully relax to zero in a \emph{zero-temperature} energy
minimization context as is adopted here.

Alternatively, as shown in the finite element (FE) context
\citep{Feng2006}, tangential vertex motions may be suppressed
partially by enforcing the incompressibility of the membrane as a
local (rather than global) area constraint.  This approach also allows
for diffusion of vertices; however, local enforcement of
incompressibility can lead to severe distortion of elements in the
mesh, and therefore hinders the simulation of large vesicle
deformations.  Furthermore, local incompressibility does not
completely suppress spurious modes, and though these degeneracies do
not prevent simulation of unforced vesicle equilibrium, they can lead
to catastrophic numerical instabilities when externally applied forces
are introduced.

Notably, these issues may be avoided entirely through the development
of meshless numerical methods for membrane mechanics.  Examples
include, Ritz methods with global basis functions (e.g., spherical
harmonics) \cite{Canham1970, Heinrich1993, Heinrich1999}, phase-field
methods \cite{Du2004, Du2006}, and moving-least squares approximation
\citep{NoguchiGompper2006}.  Yet, these approaches are not without
their own limitations (e.g., aliasing, difficulty with application of
external forces).

In this paper we propose a viscous regularization technique to
stabilize tangential motions of nodes in a FE membrane
model while enforcing incompressibility as a global constraint.  We
demonstrate the computational efficiency and effectiveness of this
approach by comparing simulation times with and without regularization
for shape transitions previously computed in \citep{Feng2006}.
Secondly, we examine the efficiency gained by enforcing the
\emph{global} constraints on membrane area and volume with an
augmented Lagrangian approach instead of the previous penalty
approach.  Lastly, we apply the regularization and constraint methods
to the simulation of two membrane shape change problems involving large
deformations and the application of external forces.  The first of
these problems is the simulation of the tether instability in a
vesicle under tension between two opposing point forces; the second is
the simulation of separation and domain formation in a two-lipid-phase
vesicle.  In the latter we demonstrate how the viscous regularization
technique can be slightly modified to formulate an $r$-adaptive
remeshing method, wherein nodes ``flow'' on the surface of the vesicle
in such a way as to avoid element distortion.

The outline of this paper is as follows: Section 2 briefly introduces
the FEM formulation for bilayer membrane mechanics along with
artificial viscosity mesh stabilization and augmented Lagrangian
constraint enforcement; Section 3 shows two applications: tether
formation and lipid phase separation, based on the methods described
in Section 2; Section 4 concludes with discussions of results and
future applications.
\section{Methods}\label{sec:methods}

\subsection{Lipid bilayer mechanics and finite element approximation}\label{sec:bilayers}

We begin with a brief review of the mechanics of bilayer membranes and FEM
approximation we use. For further details, the reader is refer to the
paper \cite{Feng2006}.
  
\begin{figure}[ht]
\centering
\includegraphics[width=0.5\textwidth]{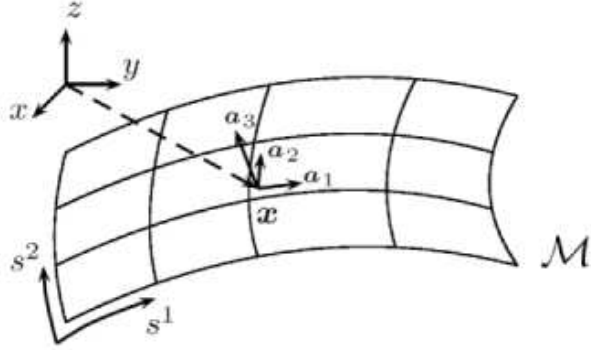}
\caption{\label{fig:surface}Geometry of a surface} 
\end{figure}
\paragraph*{Membrane kinematics.}
The bilayer membrane is described as a two-dimensional surface
$\mathcal{M}$ embedded in three-dimensional space
(Fig.\ref{fig:surface}), parameterized by curvilinear coordinates
$\{s^1,s^2\}$, such that its position is given by the map
$\bm{x}:\mathbb{R}^2 \to \mathbb{R}^3$.  
With standard definitions from differential geometry
\citep{Sokolnikoff,DoCarmo} we can span the surface tangent plane with both
(covariant) basis vectors $\bm{a}_\alpha =
\frac{\partial\bm{x}}{\partial s^\alpha} \equiv \bm{x}_{,\alpha}$ and
dual (contravariant) basis vectors $\bm{a}^\alpha$ defined such that
$\bm{a}^\alpha\cdot\bm{a}_\beta = \delta^\alpha_\beta$.
The covariant and contravariant surface metric tensors are then 
\begin{equation}
a_{\alpha \beta} = \bm{a}_\alpha \cdot \bm{a}_\beta ,
 \quad \text{and} \quad 
a^{\alpha \beta} = \bm{a}^\alpha \cdot \bm{a}^\beta ,
\end{equation}
and the determinant of the covariant metric tensor will be denoted
\begin{equation}
a = \det a_{\alpha\beta} .
\end{equation}
The normal to the surface is
\begin{equation}
\bm{d} \equiv \bm{a}_3 = \frac{\bm{a}_1 \times \bm{a}_2}{|\bm{a}_1 \times \bm{a}_2|} = \frac{\bm{a}_1
\times \bm{a}_2}{\sqrt{a}}.
\end{equation}
The curvature tensor $\mathbf{B}$ is defined by its covariant components
\begin{equation}
b_{\alpha \beta} = - \bm{d}_{,\alpha} \cdot \bm{a}_\beta 
= \bm{d} \cdot \bm{x}_{,\alpha \beta} 
= \bm{d} \cdot \bm{a}_{\alpha,\beta} .
\end{equation}
The mean curvature is one half of the trace of the curvature tensor
\begin{equation}
H = \frac{1}{2} b^\alpha_\alpha = \frac{1}{2} a^{\alpha\beta}b_{\alpha\beta}
= -\frac{1}{2} a^{\alpha\beta}(\bm{d}_{,\alpha}\cdot\bm{a}_\beta)
= -\frac{1}{2} \bm{a}^\alpha \cdot \bm{d}_{,\alpha}
\end{equation}
(where $a^{\alpha\beta}$ is the contravariant metric tensor,
defined such that
$a^{\alpha\beta}a_{\alpha\beta}=\delta^\alpha_\beta$), and the
Gaussian curvature is the determinant of the curvature tensor
\begin{equation}
K = \det \mathbf{B} = \det b_{\alpha\beta} .
\end{equation}

\paragraph*{Lipid bilayer mechanics.}We describe the energetics of the
membrane by the Helfrich model \cite{Helfrich1973}, which assumes a
strain energy of the form
\begin{equation}
E[\bm{x}] = 
\int_\mathcal{M} \frac{1}{2}\mathcal{K}_C (2 H - C_0)^2 \sqrt{a} d^2s +
\int_\mathcal{M} \mathcal{K}_G K \sqrt{a}  d^2s
\end{equation}
where $\mathcal{K}_C$ is the bending modulus and $\mathcal{K}_G$ is
the Gaussian curvature modulus. By the Gauss-Bonnet theorem
\cite{Sokolnikoff}, the integral of Gaussian curvature is a
topological constant $\int_{\mathcal{K}_G}K\sqrt{a}d^2s=4\pi(1-g)$,
with $g$ being the genus, i.e. the number of handles, and thus can be
neglected.

The weak form of equilibrium for the membrane can be obtained in
general by the principle of virtual work, and for the case of
conservative loads by minimization of total potential energy.  The
later dictates that the total potential energy by stationary with
respect to any arbitrary admissible surface variation $\delta\bm{x}$
\begin{equation}
\delta \Pi \equiv \delta E - \delta W^\text{ext} = 0.
\end{equation}
Here $\delta E$ is the first variation of the  membrane bending energy,
%
and $\delta W^\text{ext}=\int_\mathcal{M} \bm{f}^\text{ext} \cdot
\delta\bm{x} \sqrt{a}$ is virtual work done by conservative external forces
$\bm{f}^\text{ext}$.
A straightforward calculation \citep{Feng2006} gives the first variation of the total energy as
%
%
\begin{equation}\label{eq:virtualWork}
\delta \Pi = \int_\mathcal{M}\left[
\bm{n}^\alpha \cdot \delta \bm{a}_\alpha + 
\bm{m}^\alpha \cdot \delta \bm{d}_{,\alpha} 
- \bm{f}^\text{ext} \cdot \delta\bm{x}  \right]\sqrt{a}{ds^1}{ds^2} ,
\end{equation}
where we have defined stress resultants $\bm{n}^\alpha$ and moment resultants $\bm{m}^{\alpha}$ as
\begin{equation}
\bm{n}^\alpha =  
\mathcal{K}_C(2H-C_0)a^{\alpha\beta}\bm{d}_{,\beta} + \mathcal{K}_C \frac{1}{2}(2H-C_0)^2  \bm{a}^\alpha 
\qquad
\bm{m}^{\alpha} = -\mathcal{K}_C (2H-C_0) \bm{a}^\alpha . 
\end{equation}
%


\paragraph{Enforcing constraints: augmented Lagrangian method.}

Admissibility requirements on trial functions $\bm{x}$ and variations
$\delta\bm{x}$ include the satisfaction of any active constraints,
such as the aforementioned constraints on total surface area and
enclosed volume.  Here we will enforce these constraints with the
augmented Lagrangian (AL) approach (see, e.g.,
\citep{NocedalWright1999}).  The AL method may be thought of as a
hybrid between penalty and Lagrange multiplier methods.  The basic
idea of AL is to solve iteratively for a Lagrange multiplier,
computing multiplier updates from a penalty term.  
To enforce constraints on both area and volume of a membrane, we
establish a sequence of modified energy funcitonals, the $n$th of
these taking the the form $I^n = \Pi + I^\text{con}$, where
$I^\text{con}$ is a constraint energy term
\[
I^\text{con} = 
\frac{\mu_V}{2}(V-\bar{V})^2 -p^{n}V + 
\frac{\mu_A}{2}(A-\bar{A})^2 + \alpha^{n}A .
\]
Here $\bar{A}$ and $\bar{V}$ are the specified surface area and
enclosed volume of the membrane, $\mu_A$ and $\mu_V$ are penalty
parameters (large and positive), and $\alpha^{n}$ and $p^{n}$ are
tension and pressure multiplier estimates for the $n$th iteration.
Minimization of the modified energy (holding multiplier estimates
fixed) yields
\[
\delta I^{n} = \delta \Pi - p^{n+1}\delta V + \alpha^{n+1} \delta A  = 0,
\]
where $p^{n+1} = p^{n} - \mu_V(V_{n}- \bar{V})$ and $\alpha^{n+1} =
\alpha^{n} + \mu_A(A_{n}- \bar{A})$ are the updated multiplier
estimates. Iteration of minimization followed by multiplier updates is
continued until constraints are satisfied to within some preselected
tolerance, TOL, as shown below in Algorithm \ref{algorithm:AL}.  In
this way the modified energy converges to the pure Lagrange-multiplier
constrained functional, with the added benefit of avoiding the
associated saddle-point problem, retaining a minimization structure
which is convenient for nonlinear optimization algorithms.


Whereas pure penalty methods require very large penalty parameters for
accurate constraint enforcement, the AL iterative updates can achieve
accuracy with much smaller penalty terms.  In practice this is an
important advantage, since when the penalty parameters $\mu_V$ and
$\mu_A$ become large, numerical minimization becomes difficult as the
Hessian $\nabla^2 E$ (or stiffness matrix) becomes quite ill
conditioned near the minimizer. This property makes minimization
algorithms like quasi-Newton and conjugated gradient perform poorly,
as finding the search directions becomes difficult
\cite{NocedalWright1999}.  However, small penalty parameters can
produce a large number of AL iterations for convergence.  Hence, it is
common in practice to incrementally increase penalty parameters by
some factor, FAC, after each AL iteration process to achieve faster
convergence.  These penalty parameter updates are also included in
Algorithm \ref{algorithm:AL}.

\begin{algorithm}
\caption{Augmented Lagrangian method to enforce global area and volume
constraints.} \label{algorithm:AL}
\begin{algorithmic}  
\STATE Initialize: $p^0$, $\alpha^0$, $\mu_{V}^0$, $\mu_{A}^0$,
\STATE Set n=0
\REPEAT
\STATE Minimize $I^n=\Pi + \frac{\mu_V}{2}(V-\bar{V})^2 -p^{n}V + \frac{\mu_A}{2}(A-\bar{A})^2 + \alpha^{n}A $
\STATE Updates: \\
\quad
$p^{n+1} = p^{n} - \mu_V^n(V_{n}- \bar{V})$ ,
\quad
$\alpha^{n+1} = \alpha^{n} + \mu_A^n(A_{n}- \bar{A})$
\\
\quad
$\mu_{V}^{n+1}=\mu_{V}^{n}\times\text{FAC}$ ,
\quad
$\mu_{A}^{n+1}=\mu_{A}^{n}\times\text{FAC}$ 
\\
\quad
$n\leftarrow n+1$
\UNTIL{\quad $|V^{n}-\bar{V}| < \text{TOL} \quad\text{and}\quad |A^{n}-\bar{A}| <
\text{TOL}$}
\end{algorithmic}
\end{algorithm}

\paragraph*{Finite element approximation.} 
A FE approximation is introduced by replacing the field $\bm{x}$ with
the approximated field $\bm{x}_h$ defined by
%
\begin{equation}
\bm{x}_h(s^1,s^2) = \sum_{a=1}^{N} \bm{x}_a N^a(s^1,s^2) 
\end{equation}
where the $N^a(s^1,s^2),~ a=1,\dots, N$ are \emph{shape functions} of
the FE mesh, and their coefficients, $\bm{x}_a$ are the
positions of the nodal control vertices.  Introducing this
approximation into the modified energy funcitonal upon minimization
leads to a set of discrete approximate equilibrium equations
\begin{equation}\label{eq:FEequilibrium}
\bm{f}^\text{int}_a + \bm{f}^\text{con}_a - \bm{f}^\text{ext}_a = 0 .
\end{equation}
Here $\bm{f}^\text{int}_a$ are the \emph{internal} nodal forces due to
bending of the membrane,
\begin{equation}\label{eq:nodalInternalForce}
\bm{f}^\text{int}_a = \int_\mathcal{M} \left[
\bm{n}^\alpha \cdot \frac{\partial\bm{a}_\alpha}{\partial\bm{x}_a} +
\bm{m}^\alpha \cdot \left(\frac{\partial\bm{d}}{\partial\bm{x}_a}\right)_{,\alpha} 
\right] \sqrt{a} {d^2s} ;
\end{equation}
$\bm{f}^\text{con}_a$ are the \emph{constraint} nodal forces due to
the pressure and tension that are conjugate to the constrained volume
and area,
\begin{equation}\label{eq:nodalConstraintForce}
\bm{f}^\text{con}_a = 
-p^{n+1}\frac{\partial V}{\partial\bm{x}_a} + 
\alpha^{n+1}\frac{\partial A}{\partial\bm{x}_a} ;
\end{equation}
and $\bm{f}^\text{ext}_a$ are the \emph{external} nodal forces, due to
the application of distributed loads on the surface,
\begin{equation}\label{eq:nodalExternalForce}
\bm{f}^\text{ext}_a = \int_\mathcal{M} \bm{f}^\text{ext} N^a \sqrt{a}
{d^2s} .
\end{equation}
Note that the integrands of the global expressions for internal and
constraint forces are described in more explicit detail in
\citep{Feng2006}.  Following that work, we again employ
$C^1$-conforming subdivision surface shape functions
\cite{CirakOrtiz2000, CirakOrtiz2001} along with second-order
(three-point) Gaussian quadrature for the computation of element
integrals.

\subsection{Viscous regularization of tangential mesh deformation}\label{sec:stabilization}

In the curvature model, the energy is determined by the mean curvature
which is a parameterization-independent property of the surface shape,
and thus is not sensitive to in-plane dilatational or shearing
deformations of the surface FE mesh.  Much like physical lipid
molecules, FE nodes can flow freely on the deformed surface. As
discussed in \cite{Feng2006}, this fact is manifested in the
appearance of degenerate, zero-stiffness, zero-energy modes. Here we
discuss the implementation of an artificial viscosity method designed
to numerically eliminate these degenerate modes.

For solid shells having both reference and deformed configurations,
in-plane deformations (dilatation and shearing) can thus be expressed
locally in terms of first derivatives of the surface position maps of
these two configurations. In curvature model,a well-defined reference
configuration does not exist since the energy is only related to the
deformed shape. The basic ingredients for stabilization of these
tangential modes are the introduction of a reference configuration and
an energy term elastically penalizing in-plane deformation away from
this reference state.  However, to retain the physics of the original
model, the addition of any in-plane elastic energy must result in a
variational problem possessing the same minimizing solution as the
original problem.  In other words, the artificial in-plane energy must
attain a value of zero when the entire model is in equilibrium.  To
design an algorithm that achieves these goals, we define a sequence of
variational problems, minimizing a modified energy functional
\begin{equation}\label{eq:regularizedEnergy}
I^n = \Pi[\bm{x}] + I^\text{con}[\bm{x}] + I^\text{reg}[\bm{x}; \bm{X}^n]
\end{equation}
where the reference configuration $\bm{X}^n$ for the $n$th iteration
is the deformed solution $\bm{x}^{n-1}$ of the previous iteration.
The form of the regularization energy $I^\text{reg}[\bm{x}; \bm{X}]$
can be chosen such that it vanishes when $\bm{x}=\bm{X}$, to ensure that
solutions $\bm{x}^n$ converge to minimizers of the original
unregularized problem with increasing $n$.  This regularization method
is outlined below in Algorithm \ref{algorithm:reg}.
%

\begin{algorithm}
\caption{\label{algorithm:reg}Viscous regularization via reference updates.} 
\begin{algorithmic}  
\STATE Set $X^0=$ initial shape.
\STATE Set $n=0$
\REPEAT
\STATE Minimize $I^n = \Pi[\bm{x}] + I^\text{con}[\bm{x}] + I^\text{reg}[\bm{x}; \bm{X}^n]\quad \rightarrow \quad \text{solution, }\bm{x}^n$
\STATE Update reference: set $\bm{X}^{n+1}=\bm{x}^n$
\STATE $n\leftarrow n+1$
\UNTIL{$I^\text{reg}[\bm{x}^n;\bm{X}^{n}] < \text{TOL}$}  
\end{algorithmic}
\end{algorithm}

Qualitatively, assignment of the reference configuration for each
iteration to be the current configuration of the previous iteration
results in a type of algorithmic viscosity, producing forces that
resist the motion of nodes away from their position at each previous
iteration.  The quantitative details of this viscosity depend on the
particular form chosen for the in-plane regularization energy,
$I^\text{reg}$.  Here we give two example forms, the first derived from
planar continuum elasticity theory and the second representing the mesh as a
network of viscous dashpot elements.


\paragraph{Continuum elastic regularization energy.} 
Here we treat the in-plane deformation response for each
regularization iteration as that of a two-dimensional solid membrane.
This local response can be modeled via a hyperelastic strain energy density,
$w(\mathbf{F})$, which is a function of the surface deformation
gradient
\begin{equation}
\mathbf{F} = \bm{a}_\alpha \otimes \bm{A}^\alpha ,
\end{equation}
where $\bm{A}^\alpha$ are the dual basis vectors on the reference
surface, i.e., $\bm{A}^\alpha\cdot\bm{A}_\beta =
\delta^\alpha_\beta$, where $\bm{A}_\alpha = \bm{X}^n_{,\alpha}$.  Thus
the regularization energy becomes
\begin{equation}
I^\text{reg}[\bm{x};\bm{X}^n] = \int_\mathcal{M} w(\mathbf{F}) \sqrt{A} d^2s .
\end{equation} 

To preserve objectivity, the strain energy is a function of
$\mathbf{F}$ through implicit dependence on the invariants of the
surface-Right-Cauchy-Green deformation tensor $\mathbf{C} =
\mathbf{F}^\textsf{T}\cdot\mathbf{F} =
a_{\alpha\beta}\bm{A}^\alpha\otimes\bm{A}^\beta$ \cite{Steigmann}.  As
$\mathbf{C}$ is a rank-2 tensor, the two non-zero principal invariants
are
\begin{subequations}
\begin{align}
I_1 &= \text{tr}(\mathbf{C}) = \bar{a}^{\alpha\beta}a_{\alpha\beta} \\
I_2 &= \frac{1}{2}\{[\text{tr}(\mathbf{C})]^2 - \text{tr}(\mathbf{C}^2)\} 
=\frac{1}{2}\{ 
(\bar{a}^{\alpha\beta}a_{\alpha\beta})^2
- \bar{a}^{\alpha\mu}\bar{a}^{\beta\nu} a_{\alpha\beta}a_{\mu\nu}
 \}
\equiv J^2.
\end{align}
\end{subequations}
The strain energy density is thus a function of these two invariants
\[
w(\mathbf{F}) = w(I_1,I_2).
\]
As a specific example, consider a strain energy function that
decouples the dilatational, and shear responses, as used by
Evans and Skalak \cite{Evans1980} to model the red blood cell
cytoskeleton\[ w = 
\underbrace{\frac{k}{2}(J-1)^2}_\text{area change} +
\underbrace{\mu\left(\frac{\text{tr}(\mathbf{C})}{2J}-1\right)}_\text{shear} .
\]
Here $k$ and $\mu$ are stretching and shear moduli, respectively.
It should be carefully noted that although we follow here the
formalism of solid mechanics, the reference configuration $\bm{X}^n$
is not permanent as for a solid; rather the reference configuration is
iteratively updated so that the resulting stresses may relax to zero.

\paragraph{Dashpot regularization energy.}
The viscous character of our proposed scheme is much more obvious when
we compose the regularization energy of contributions from Hookean springs
placed along all element edges, namely,
\begin{equation}\label{eq:dashpotEnergy}
I^\text{reg} = \sum_{\text{edge }ab} \frac{k}{2} (\ell_{ab}-L_{ab})^2 ,
\end{equation}
where $\ell_{ab}=|\bm{x}_a-\bm{x}_b|$ and
$L_{ab}=|\bm{X}_a^n-\bm{X}_b^n|$ are the lengths of the edge
connecting mesh vertices $a$ and $b$ in the deformed and current
configurations, respectively.  Differentiating this energy, the
corresponding force on a node $a$ from the spring connecting it along
an edge to node $b$ can be obtained as
\[
\bm{f}_{ab} = k(\ell_{ab}-L_{ab}) \bm{n}_{ab}
\]
where $\bm{n}_{ab}$ is the unit vector pointing from node $a$ to node
$b$.  Recalling that the reference configuration for the $n$th
iteration is the same as the deformed configuration of the $n-1$th
iteration, the magnitude of this force can also be written as
\[
f_{ab} = k(\ell_{ab}^n - \ell_{ab}^{n-1}) .
\]
This is easily identified as the backward-Euler time-discretization of the
force-velocity relation for a viscous dashpot
\[
f_{ab} = k\frac{\partial\ell_{ab}}{\partial t} ,
\]  
Thus, iterative reference updates of the form $\bm{X}^n =
\bm{x}^{n-1}$ have the effect of converting a network of springs into
a network of dashpots, clearly revealing the viscous character of the
regularization scheme.

We have numerically implemented both the continuum elastic and dashpot
regularization described here, and although both forms are effective
in practice we have preferred the dashpot approach for its
simplicity, efficiency, and robustness.  The remainder of the paper
focuses on the use of this second approach, demonstrating its
effectiveness in application.

\section{Applications}\label{sec:applications}

\subsection{Shape vs. reduced volume}
Even in the absence of any externally applied loads, the two
constraints on area and volume cause vesicles to transition among a
variety of interesting equilibrium shapes.
Here some of the calculations performed in \cite{Feng2006} of the
equilibrium shapes for different reduced volumes are repeated, as a
first demonstration of the effectiveness of viscous regularization.

Reduced volume $\nu$ is a geometrical quantity defined as
\begin{equation}
\nu =
\frac{V}{(4\pi/3)R{_0}^3},
\end{equation}
where $R_0=\sqrt{A/4\pi}$ is the radius of a sphere with the area $A$
of the vesicle. Reduced volume is then written as
\begin{equation}
\nu =
\frac{6\sqrt{\pi}V}{A^{3/2}}.
\end{equation}
The reduced volume is the ratio of the current volume of the vesicle
and the maximum volume that the current total area of vesicle can
ensphere. For a spherical vesicle, the reduced volume $\nu = 1$; a
vesicle of any other shape has $0<\nu < 1$.

To compute the following results, the spontaneous curvature model is
used with $C_0=0$. The modified energy is computed with loop
subdivision shell elements and second-order (three-point) Gaussian
quadrature, and minimized with the quasi-newton L-BFGS-B solver
\cite{Byrd1995, Zhu1997, lbfgsbcode}.

\paragraph{Viscous regularization.}
As a first assessment of the benefit of regularization, results are
compared with the simulations done in \cite{Feng2006}, in which local
area and global volume constraints were performed by penalty method
instead of AL method. First, the same calculation of \cite{Feng2006}
is repeated; then the viscous regularization is added, with same kind
of constraints (local area and global volume constraint) and penalty
parameters ($\mu_A=10^4R^2/\mathcal{K}_C$ for local area constraint
and $\mu_A=5\times10^4R^2/\mathcal{K}_C$ for global volume
constraint).

\begin{figure}[ht]
  \centering 
  \subfigure[Initial shape, $\nu=0.914$ (not in equilibrium)]{
    \includegraphics[width=0.25\textwidth]{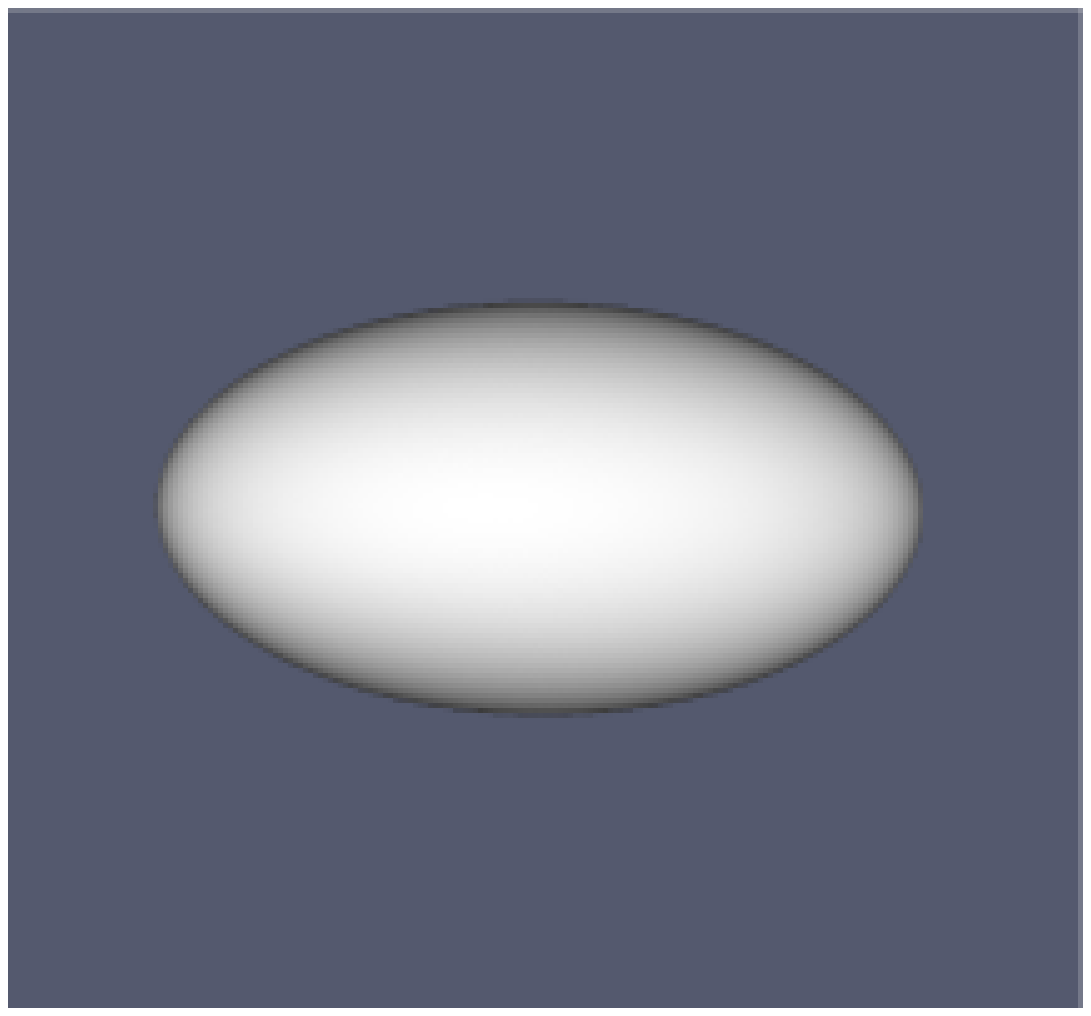}
    \includegraphics[width=0.25\textwidth]{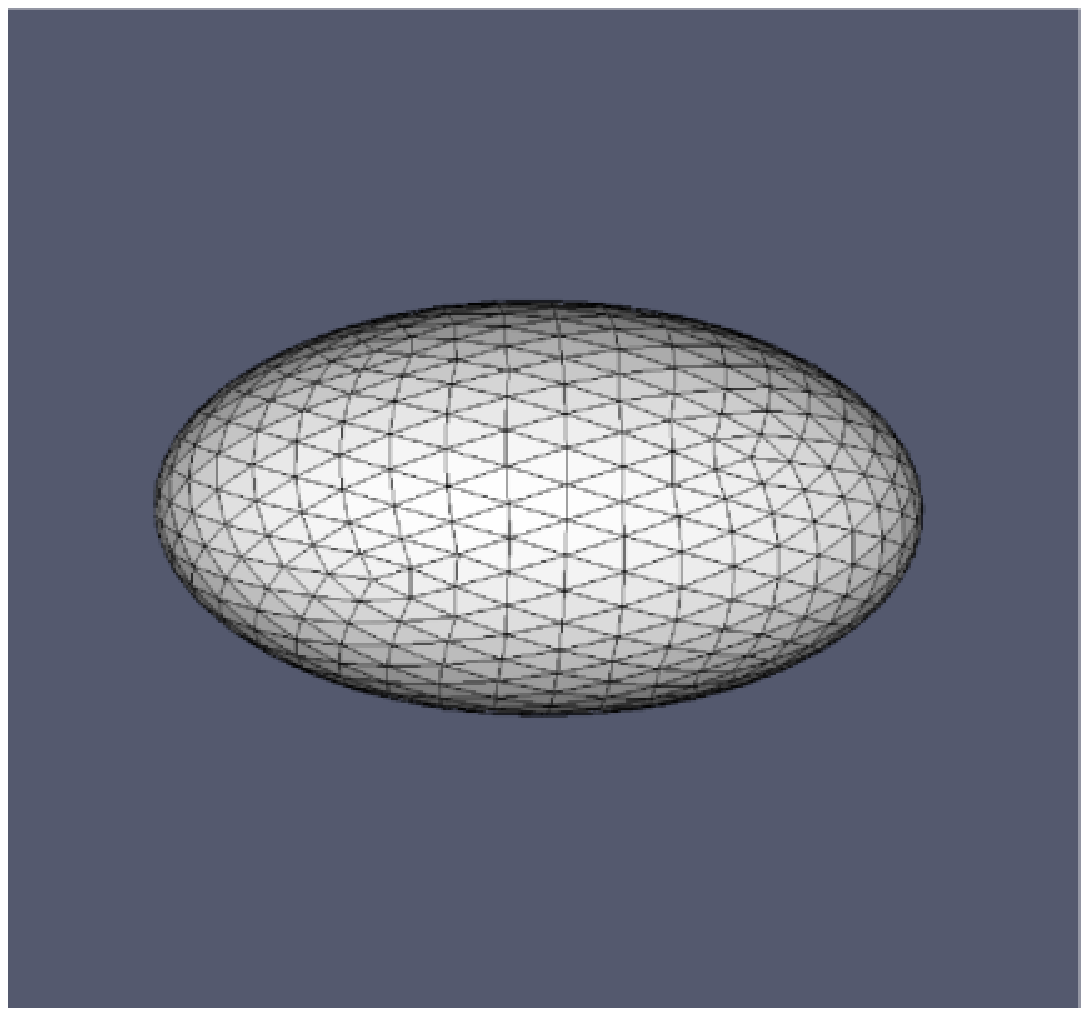}
  }
  \subfigure[$\nu=0.9$ without regularization.]{
    \includegraphics[width=0.25\textwidth]{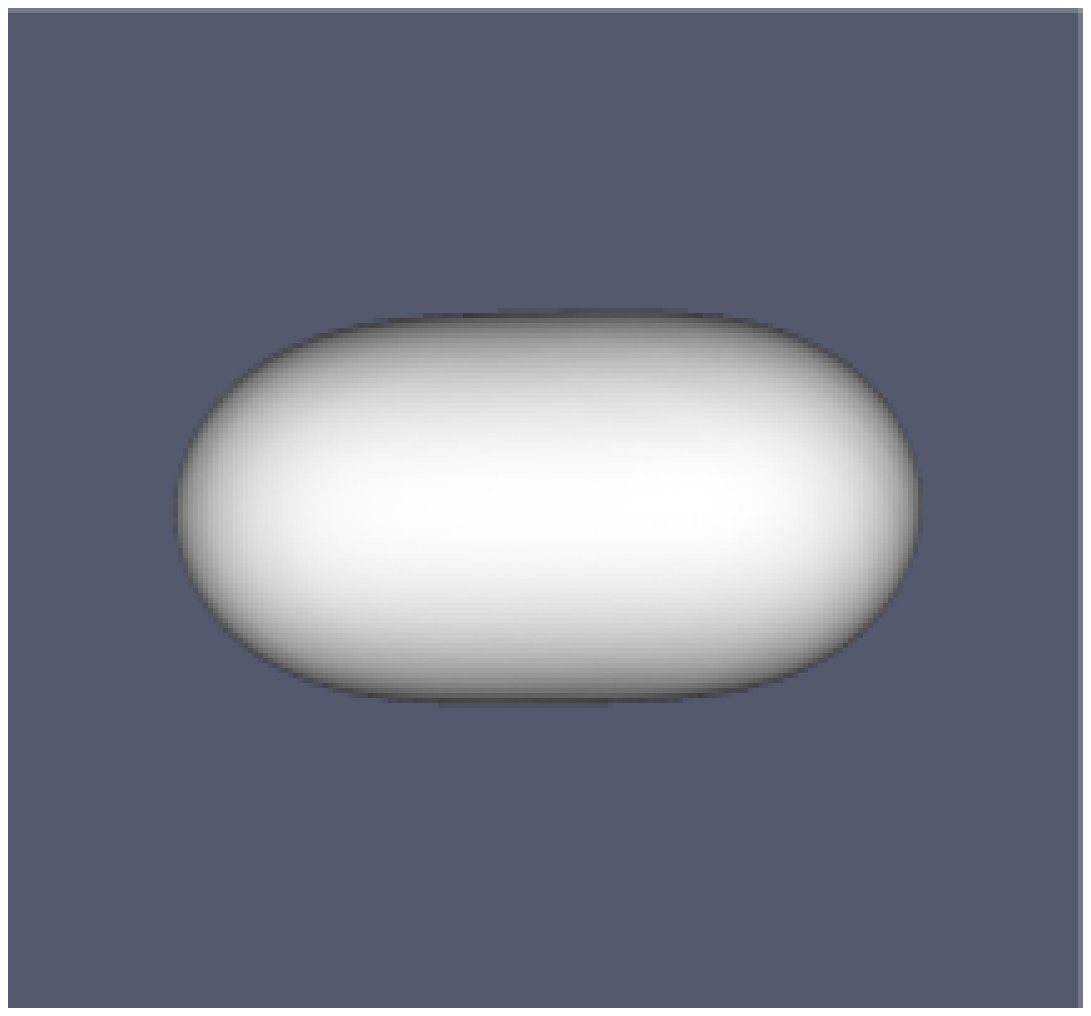}
    \includegraphics[width=0.25\textwidth]{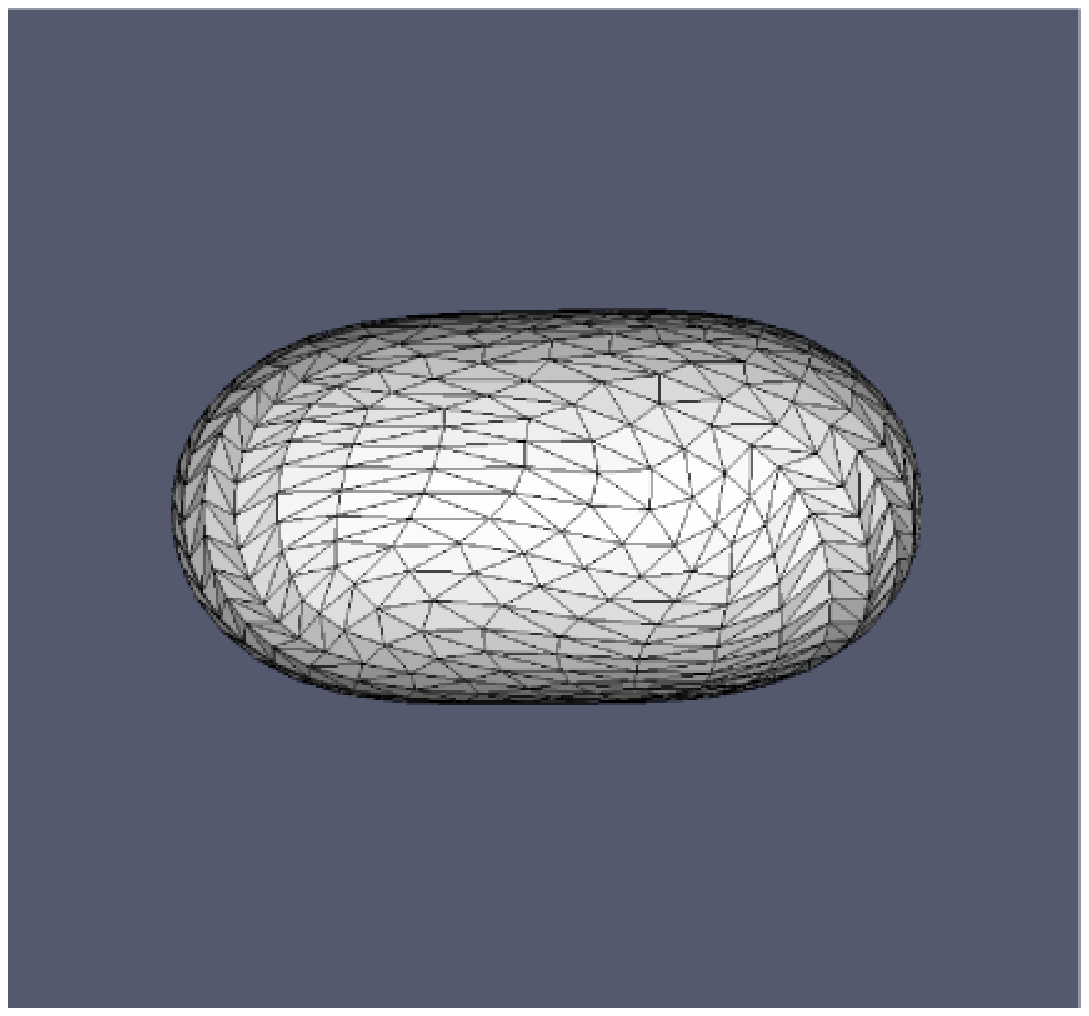}
  }
  \subfigure[$\nu=0.9$ with regularization.]{
    \includegraphics[width=0.25\textwidth]{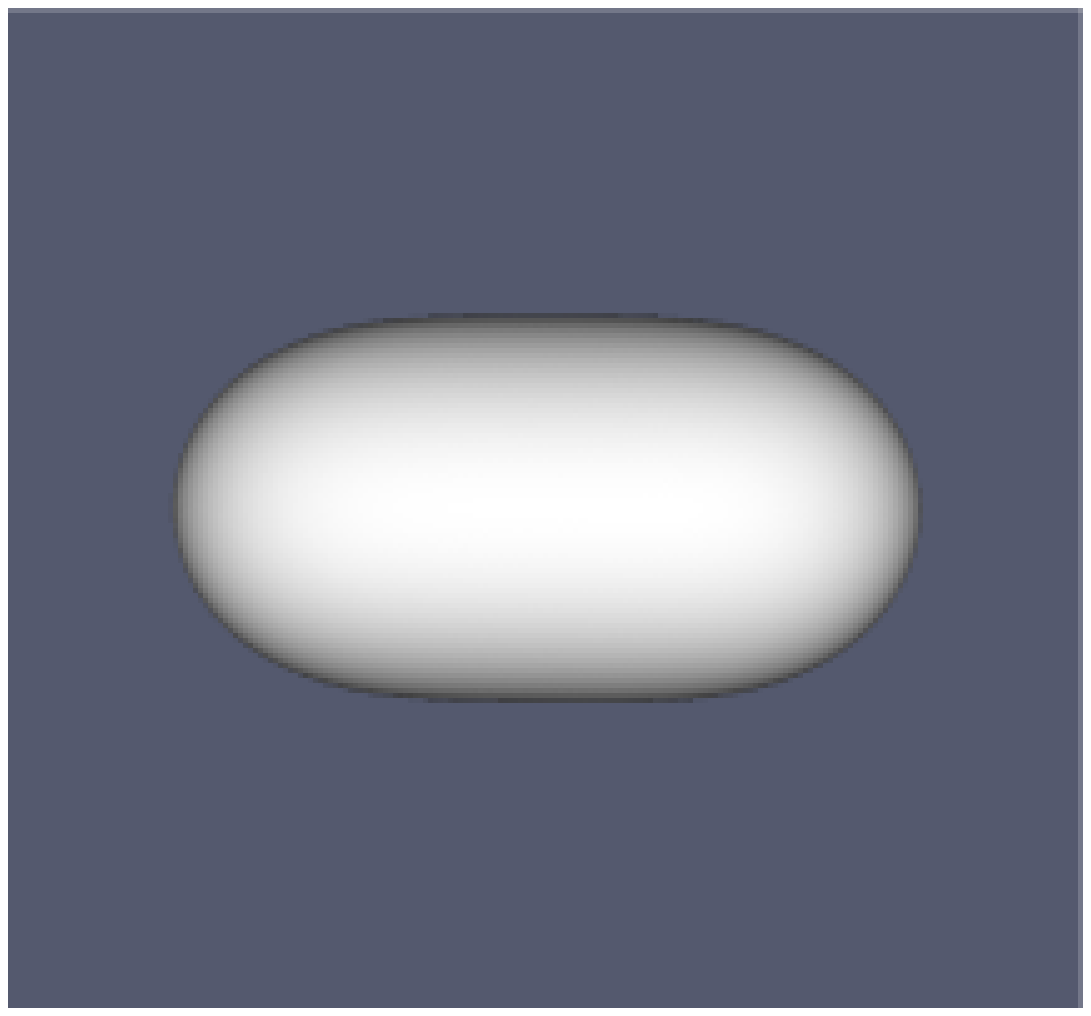}
    \includegraphics[width=0.25\textwidth]{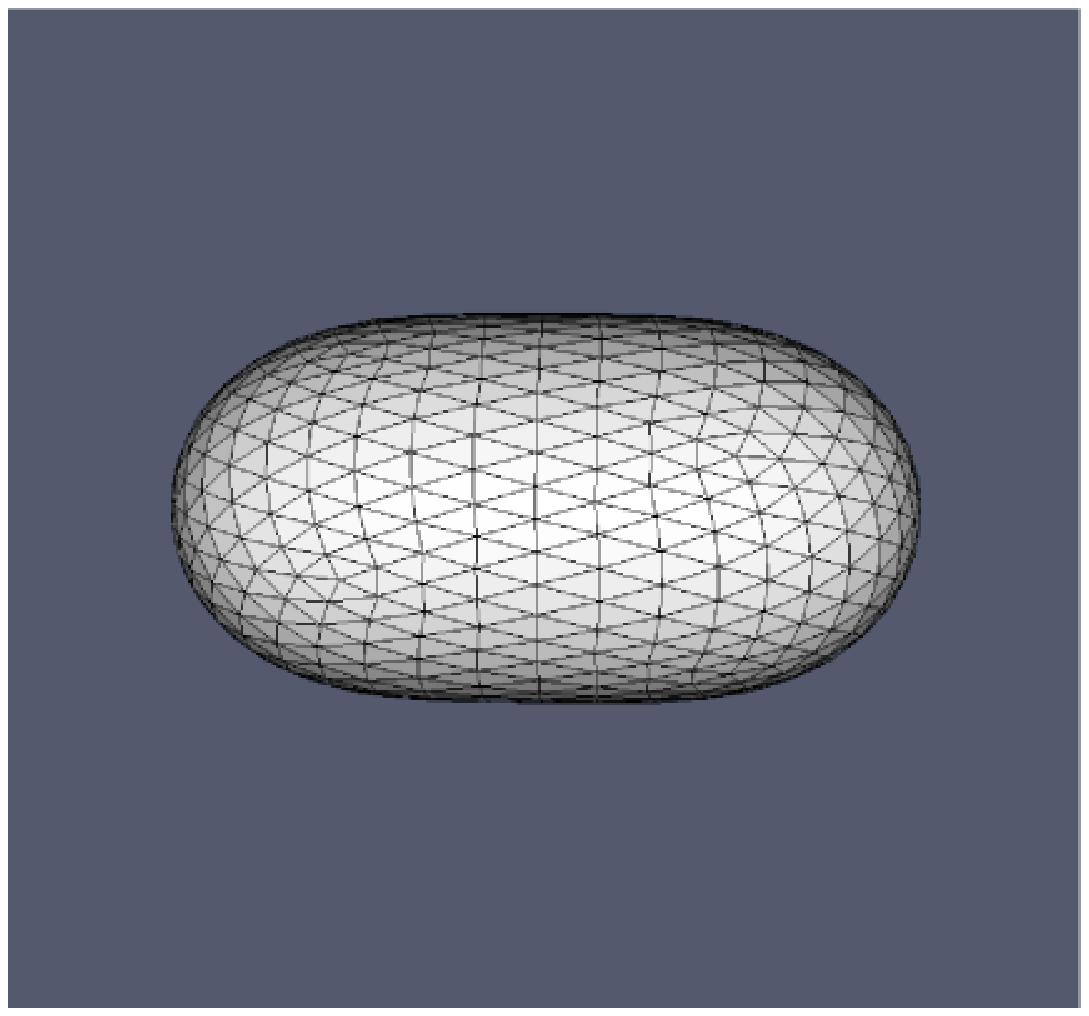}
  }\\
  \subfigure[$\nu=0.8$ without regularization.]{
    \includegraphics[width=0.25\textwidth]{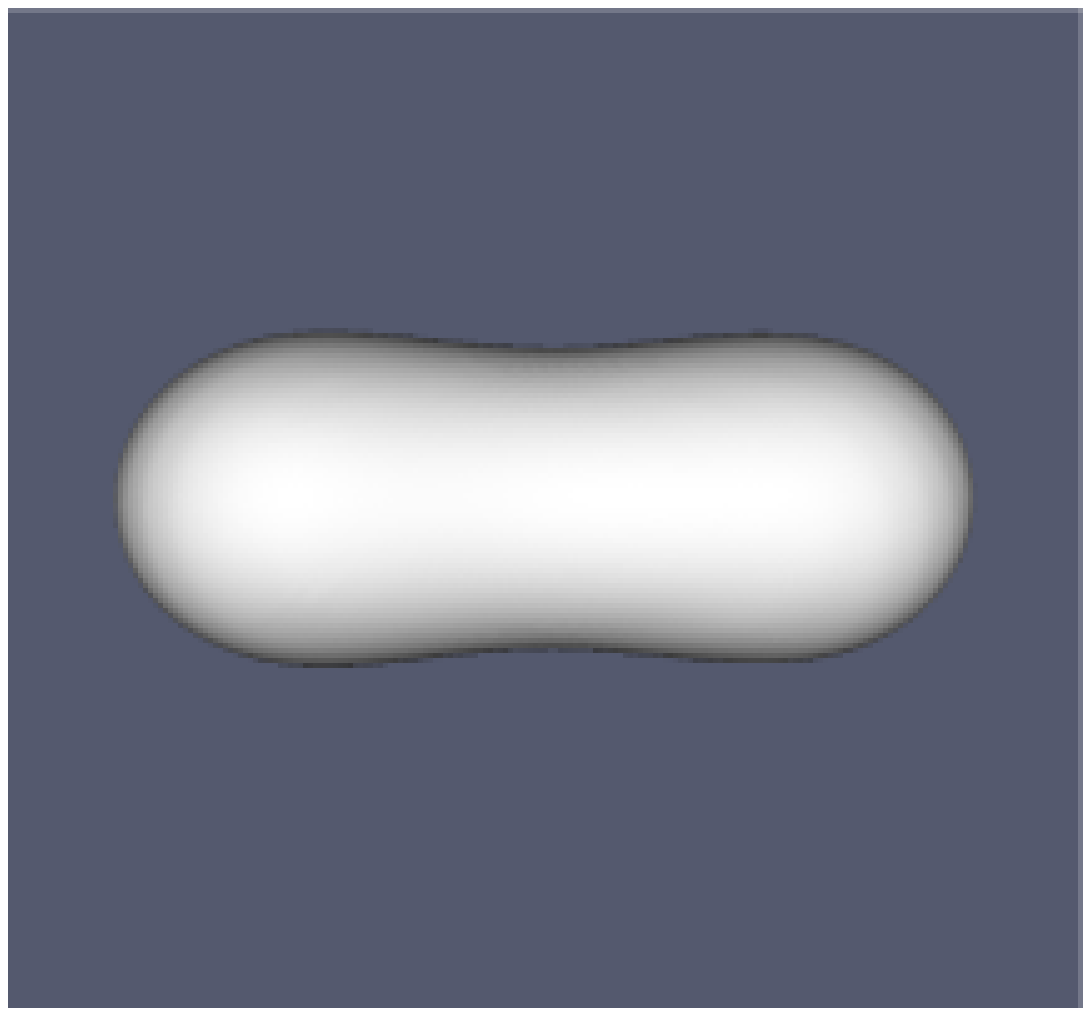}
    \includegraphics[width=0.25\textwidth]{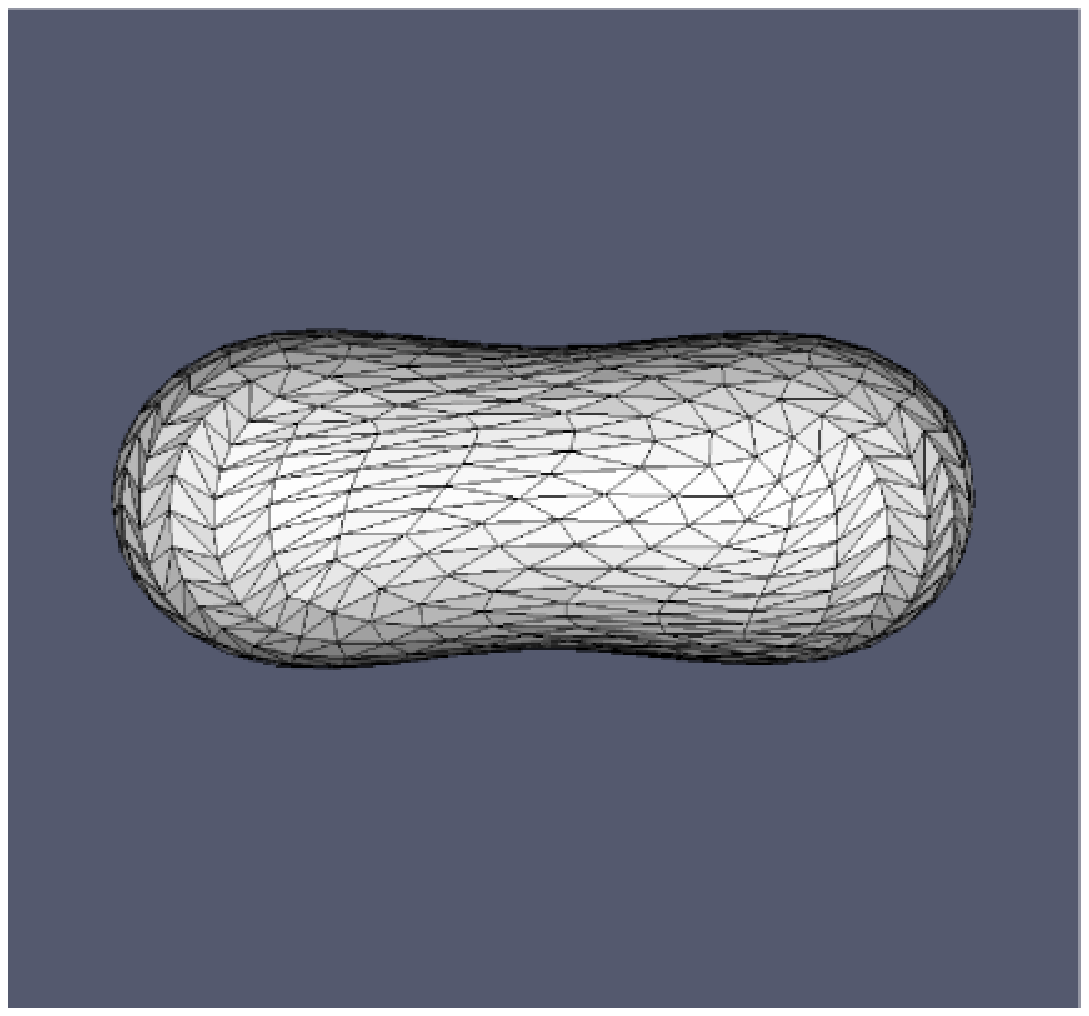}
  }
  \subfigure[$\nu=0.9$ with regularization.]{
    \includegraphics[width=0.25\textwidth]{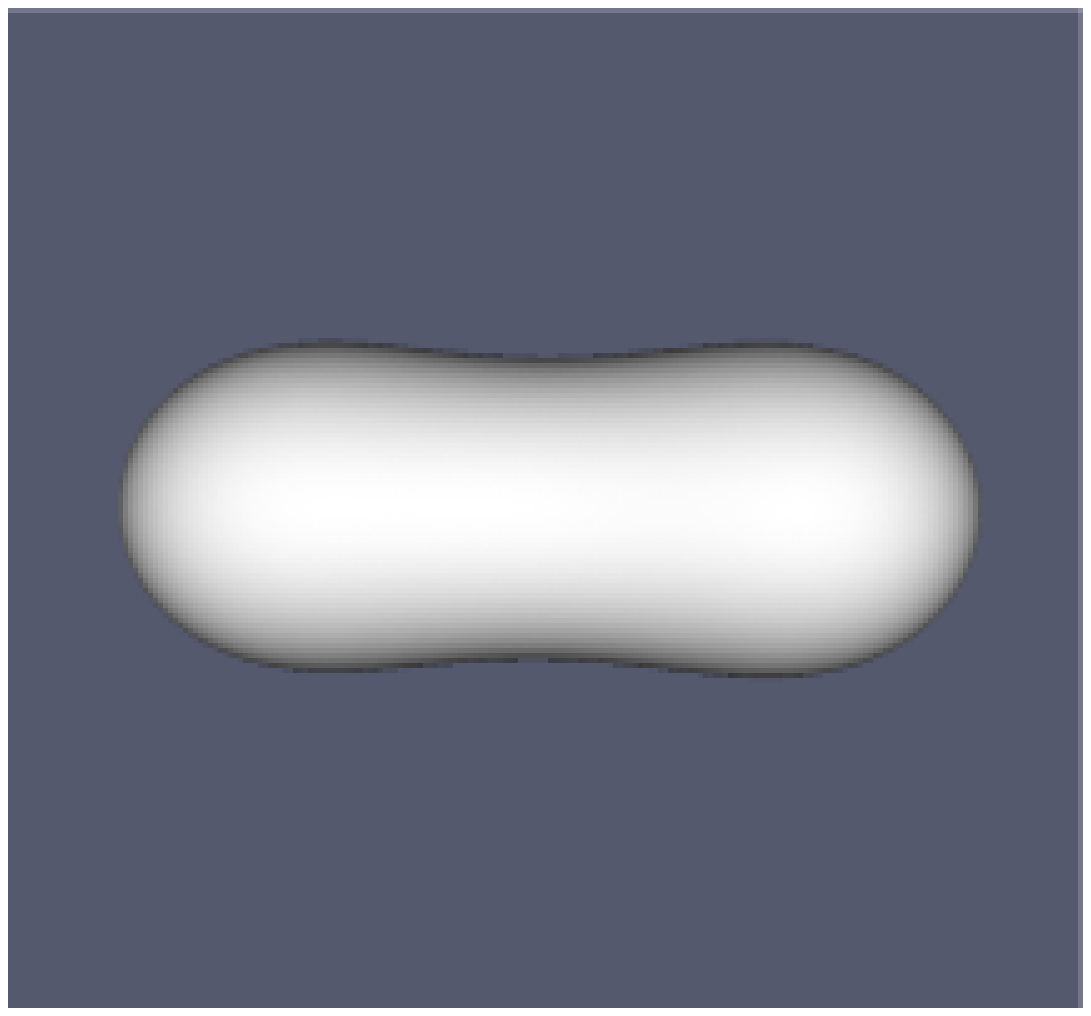}
    \includegraphics[width=0.25\textwidth]{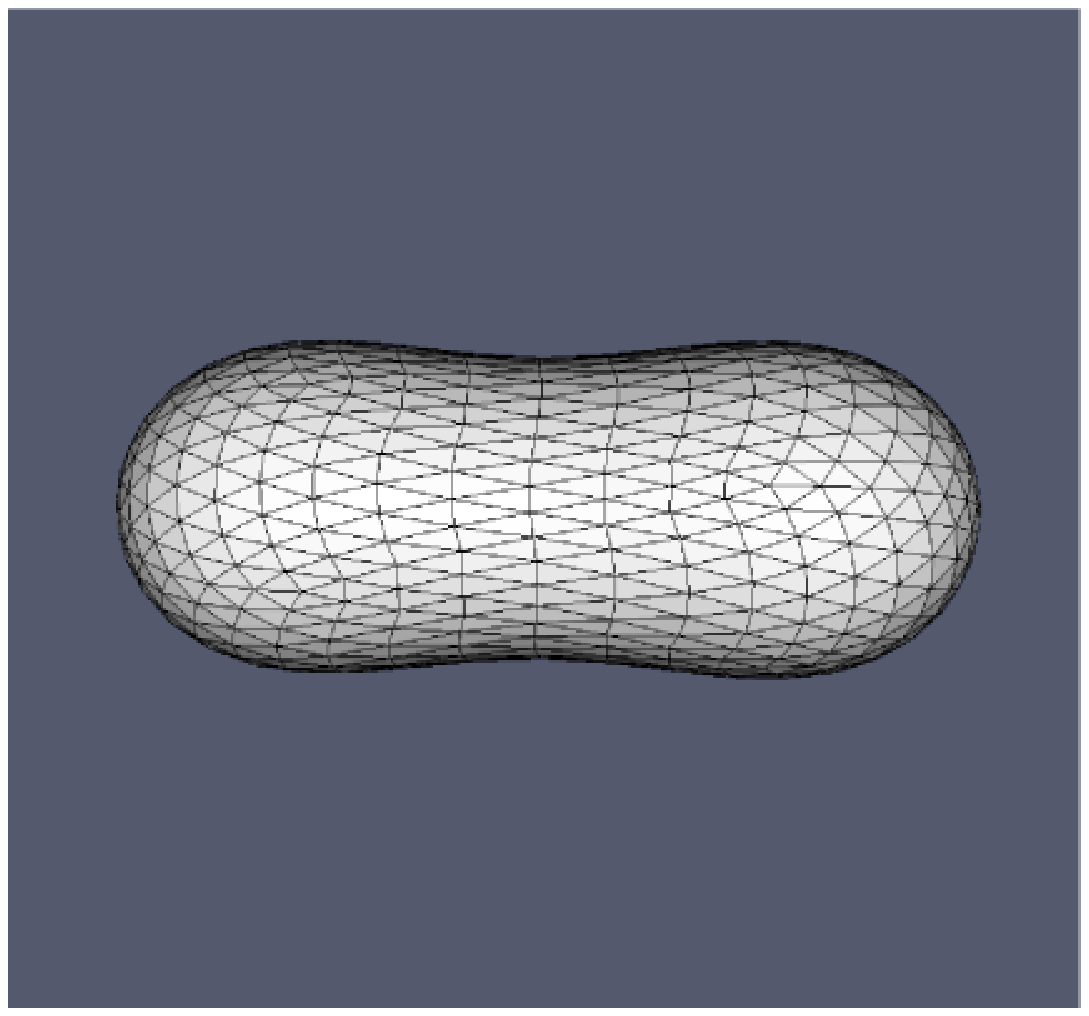}
  }
  \caption{\label{fig:nu0.90.8}Limit surfaces and
    control meshes of equilibrium shapes for $\nu=0.9$ and $\nu=0.8$.}
\end{figure}

The calculation starts from an initial ellipsoid shape which has a
reduced volume $\nu=0.914$ (Fig. \ref{fig:nu0.90.8}). In the
calculation, the area is fixed at its initial value and the volume is
reduced in order to satisfy the constraint on $\nu$. For each
simulation, violation of the volume constraint subjects the vesicle to
a large pressure according to the penalty term in the functional. The
energy is then relaxed by L-BFGS-B minimization and result in the
equilibrium shapes. The iteration of of reference updates in Algorithm
\ref{algorithm:reg} is continued until the regularization energy is
sufficiently small, $I^\text{reg}/I < 10^{-5}$.  In all the
simulations, the same mesh, made up of 642 vertex nodes and 1280
elements, is used.

The resulting equilibrium shapes for $\nu=0.9$ and $\nu=0.8$ are shown
in Fig. \ref{fig:nu0.90.8}. Starting from the initial shape, the
equilibrium shape for $\nu=0.9$ is computed by minimizing the energy;
then from the resulting $\nu=0.9$ shape, setting $\nu=0.8$, the
equilibrium shape for $\nu=0.8$ is computed. The computational cost
with and without the viscous regularization is listed in Table
\ref{table:reducedVolume}. As can be seen, the convergence rate is
highly improved (almost two orders of magnitude faster) with the
viscous regularization while the resulting shapes are equivalent. For
different choices of spring constant $k$, the computational cost also
varies. The computational cost has two contributions: one is the
iteration number for each minimization; the other is the number of
reference updates required to satisfy the convergence criterion
$I^\text{reg}/I < 10^{-5}$. These both depend on $k$. For each
minimization, the larger $k$ is, the smaller the iteration number will
be. While for the number of reference updates, it is opposite: the
larger $k$ is, the more reference updates needed. For example, to get
the equilibrium shape $\nu=0.9$ from the initial shape,
$kR^2/\mathcal{K}_C$=1 requires 2 reference updates, each of which
costs $\approx 1000$ iterations for minimization; while for
$kR^2/\mathcal{K}_C$=100, there are 20 reference updates each costing
$\approx 250$ minimization iterations. In this case
$kR^2/\mathcal{K}_C$=1 works the best, but the optimal $k$ depends on
the specific problem. In the later sections on tether formation, a
much larger $k$ ($kR^2/\mathcal{K}_C$=1000) is used.


\begin{table}
\centering
\begin{tabular}{c|c|c}           \hline
\multicolumn{3}{c}{Computational cost with and without the viscous regularization}\\ \hline\hline
$kR^2/\mathcal{K}_C$          & Total iterations (initial shape $\rightarrow \nu=0.9$) & Total iterations ($\nu=0.9 \rightarrow 0.8$)       \\  \hline
0                             & 35,950  & 603,858     \\  \hline
0.5                           & 2,304   & 287,825      \\  \hline
1                             & 1,821   & 9,482      \\  \hline
10                            & 2,122   & 12,480      \\  \hline
100                           & 4,776   & 79,038   
\end{tabular}
\caption{\label{table:reducedVolume} Viscous regularization improves
  the convergence rate of L-BFGS-B minimization. The first row
  $kR^2/\mathcal{K}_C=0$, shows the results without the viscous
  regularization, which are identical to the approach used in
  \cite{Feng2006}. ($k$: spring constant; $R$: average radius of the
  vesicle; $\mathcal{K}_C$: bending modulus.)}
\end{table}

\paragraph*{Augmented Lagrangian constraint enforcement.} 
From the results described above, viscous regularization is shown to
be able to heavily lower the computational cost when a penalty method
is used to enforce the constraints on area and volume. However,
regularization also eliminates the need for \emph{local} enforcement
of incompressibility.  \emph{Global} constraints on area and volume
can be easily implemented via the augmented Lagrangian (AL) method
which is more efficient than the previous penalty method. Here, the
shape change from the initial ellipsoid shape to the equilibrium shape
of $\nu=0.9$ is used to compare the penalty method with the AL
method. In this test global area and global volume constraints are
carried out first by the penalty method with a range of penalty
parameters, and secondly with the AL method.  Viscous regularization
is used for both the penalty method and the AL method.  Iteration
 of reference updates is continued until the regularization energy
is sufficiently small, $I^\text{reg}/I <2.0\times10^{-5}$.

The regularization spring constant $k$ is set to be
$k=10\mathcal{K}_C/R^2$, where $R$ is the average radius of the
vesicle and $\mathcal{K}_C$ is the bending modulus. For the AL method,
the penalty parameters are initialized to be a fairly small number
($\mu_V=10^4, \mu_A=10^4$), and are then increased by a factor of 2
for each of the following minimizations.  The minimization continues
until the constraints on area and volume are satisfied to within a
tolerance and the regularization energy is sufficiently small.
Viscous regularization reference updates are included with AL
multiplier updates in a single iteration loop.  This hybrid
regularization-AL algorithm, shown in Algorithm
\ref{algorithm:ALtest}, is a combination of separate Algorithms
\ref{algorithm:AL} and \ref{algorithm:reg}.

\begin{algorithm}
\caption{\label{algorithm:ALtest} Hybrid algorithm combining AL
constraint enforcement with viscous regularization.}
\begin{algorithmic}  
 \STATE Initialize: $p^0$, $\alpha^0$, $\mu_{V}^0$, $\mu_{A}^0$, $X^0=$ initial shape, $\bar{\nu}=0.9$ (the specified reduced volume).
 \STATE Set $n=0$ 
 \REPEAT
 \STATE Minimize  $I^n=\Pi[\bm{x}] + \frac{\mu_V}{2}(V-\bar{V})^2 -p^{n}V + \frac{\mu_A}{2}(A-\bar{A})^2 + \alpha^{n}A + I^\text{reg}[\bm{x}; \bm{X}^n]$
 \STATE $\rightarrow \quad \text{solution, }\bm{x}^n$
 \STATE Update reference: set $\bm{X}^{n+1}=\bm{x}^n$
 \IF {$|\nu_n-\bar{\nu}|/\bar{\nu} > \text{TOL}_1$} 
 \STATE AL Updates: \\
 \quad
 $p^{n+1} = p^{n} - \mu_V^n(V_{n}- \bar{V})$ ,
 \quad
 $\alpha^{n+1} = \alpha^{n} + \mu_A^n(A_{n}- \bar{A})$
 \\
 \quad
 $\mu_{V}^{n+1}=\mu_{V}^{n}\times\text{FAC}$ ,
 \quad
 $\mu_{A}^{n+1}=\mu_{A}^{n}\times\text{FAC}$ 
 \\
 \ENDIF
 \\ 
 $n\leftarrow n+1$
 \UNTIL{$|\nu_n-\bar{\nu}|/\bar{\nu} < \text{TOL}_1$ and $I^\text{reg}[\bm{x}^n;\bm{X}^{n}] < \text{TOL}_2$ }  
\end{algorithmic}
\end{algorithm}

As the Table \ref{table:AL} shows, the AL method reduces the
computational cost significantly.  This is especially true when high
accuracy of the constraints is desired, in which case the penalty
method requires extremely large parameters, which lead to conditioning
problems that impede convergence of the nonlinear solver.  Indeed, for
penalty parameter $>10^8$, L-BFGS-B iterations diverge. In contrast,
for the AL method to achieve high accuracy penalty parameters need not
be very large \cite{NocedalWright1999}.

\begin{table}
\centering
\begin{tabular}{l|c|c|c|c|c|c|c|c|c}          \hline
\multicolumn{10}{c}{Penalty method vs. AL method} \\ \hline\hline
Accuracy $|\nu-0.9|/0.9$          & $10^{-2}$  & $10^{-3}$  & $10^{-4}$  & $10^{-5}$  & $10^{-6}$  & $10^{-7}$  & $10^{-8}$  & $10^{-9}$ & $10^{-10}$ \\  \hline
Penalty parameter                 & $10^4$    & $10^5$     & $10^6$    & $10^7$     & $10^8$    & n/a        & n/a       & n/a      & n/a      \\  \hline
Iterations (penalty method)   & 379       & 436       & 1032      & 3455       & 9231     & n/a        & n/a         & n/a      & n/a       \\  \hline
Iterations (AL method)        & 366       & 425       & 500       & 515        & 616      & 743        & 916         & 981      &1201        
\end{tabular}
\caption{\label{table:AL} Computational cost of the penalty method and
  the AL method. To achieve the same accuracy, the AL method requires
  fewer total iterations compared to the penalty method.  More
  importantly, for extremely high accuracy ($<10^{-7}$), the L-BFGS-B
  minimization diverges with the penalty method, while the AL method
  still converges.}
\end{table}

\subsection{Tether formation}\label{sec:tetherFormation}
A point-force acting on lipid membranes can pull out a long narrow
tube commonly called a tether. This can be done by using micropipettes
\cite[e.g.,][]{Heinrich1999}, optical tweezers
\cite[e.g.,][]{Fygenson1997}, or even growing microtubules inside the
vesicle \cite{Fygenson1997}. The mechanical reason for formation of
tethers lies in the lack of shearing modulus for membranes. Elongating
in one direction and contracting in the other to such a spectacular
way like tethers mechanically means extremely large shear deformations
\cite{Bozic1997, Derenyi2002, Powers2002, Smith2004}.

\begin{figure}[ht]
\centering
  {
    \includegraphics[height=1.7in]{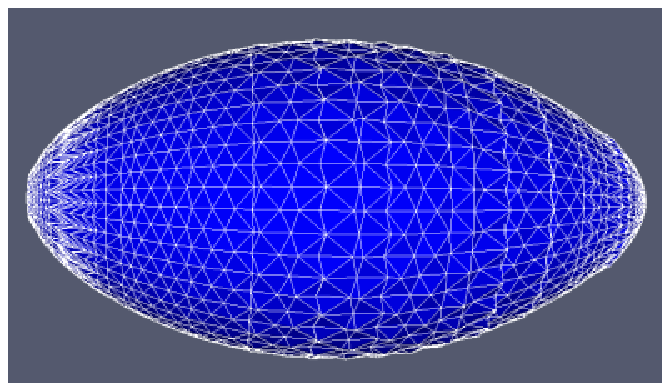}
    \includegraphics[height=1.7in]{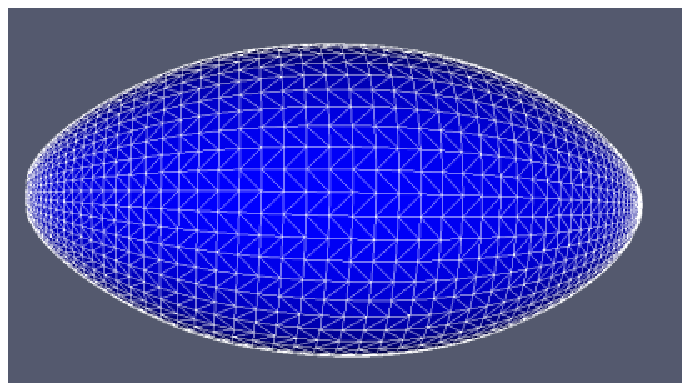}
  }
\caption{\label{fig:unstableMesh}The equilibrium shape of a vesicle of
  reduced volume $\nu=0.9$ with external forces ($\approx$ 1 pN)
  applied at the two ends (forces not shown), mesh without (left
  figure) and with (right figure) the viscous regularization.  Note
  that the unstabilized mesh is subject to element distortion even at
  small applied load.}
\end{figure}

Since tether simulation involves very large deformations, the
triangles in the finite element mesh are subject to severe
distortions. In practice, as elements become more distorted, the
zero-energy tangential modes can actually become numerically unstable
(Fig. \ref{fig:unstableMesh}). Viscous regularization has to be added
in order to suppress these zero-energy modes. Furthermore, the
critical force to pull out a tether is very sensitive to pressure and
surface tension. Numerically, this necessitates highly accurate
enforcement of the volume and area constraints. For a penalty method
this implies very large penalty parameters, which lead to conditioning
problems (e.g., Table \ref{table:AL}). For this reason, here the
augmented Lagrangian method is applied.

\begin{figure}
\centering
\includegraphics[height=5in]{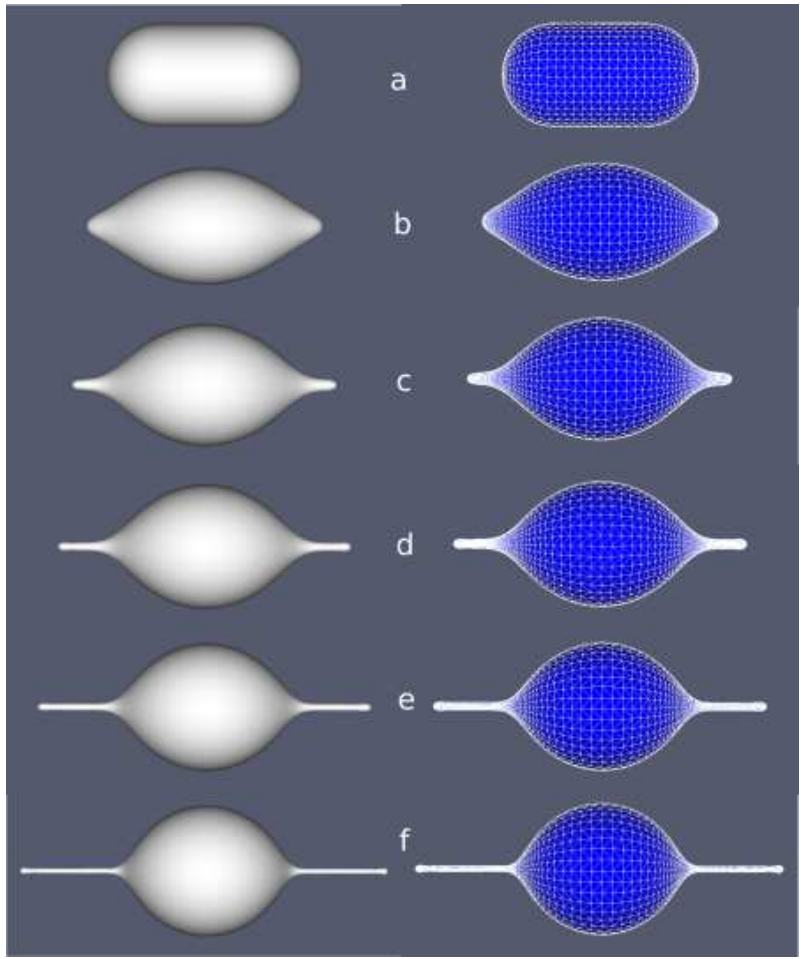} 
\caption{\label{fig:tether0.9}The tethering of a vesicle starting from
  a prolate of reduced volume $\nu=0.9$. Left: limit surfaces of
  equilibrium shapes; right: control meshes.  Number of nodes and
  elements: (a) 4202 nodes, 8400 elements; (b) \& (c) 4682
  nodes, 9360 elements; (d) 6202 nodes, 12400 elements; (e) \& (f)
  8682 nodes, 17360 elements. End to end distance:
  (a) 6.8$\mu$m, (b) 8.2$\mu$m, (c) 9.2$\mu$m, (d) 10.2$\mu$m, (e) 11.6$\mu$m, and (f) 12.8$\mu$m}
\end{figure}

Starting from an initial equilibrium shape (prolate), tether
development is simulated by incrementally displacing nodes at the tips
of the vesicle, and performing energy minimization resulting in the
equilibrium tethered shapes for each extension.  However, even with
the viscous regularization, the mesh can still be distorted by the
dramatic deformations experienced at larger extensions. Therefore,
re-meshing is performed at intervals of the extension.
Fig. \ref{fig:tether0.9} shows snapshots from a typical simulation for
reduced volume $\nu = 0.9$.

\begin{figure}[ht]
  \centering
  \includegraphics[width=0.7\textwidth]{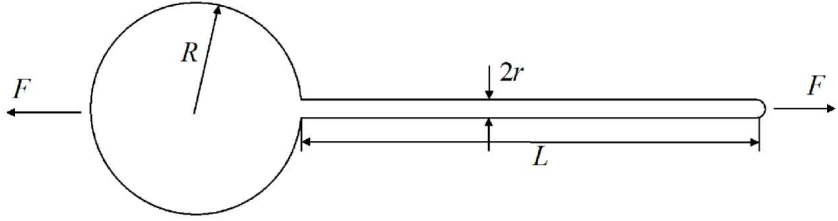}
  \caption{\label{fig:tetherShape}Schematic of the tethered shape} 
\end{figure}

\paragraph*{Applied forces.}
The reaction forces conjugate to specified end displacements can also
be calculated by simply adding up all internal forces of the fixed
nodes (Eq. \ref{eq:nodalInternalForce}). 
The force vs. end-to-end distance results for the vesicles in Fig.
\ref{fig:tether0.9} are shown in Table \ref{table:forceEst}, with $r$
the radius of the tethers ($\mu$m) and bending modulus
$\mathcal{K}_C=15K_bT$ \cite{Fygenson1997}. Although an exact
analytical solution for the force-extension relation is not possible,
a simple analytical estimate \cite{Derenyi2002} is used
to compare with the computed results from the simulation. The
estimate assumes that the thin tube (tether) is pulled out from a
sphere, and the sphere remain unchanged during the pulling (Fig.
\ref{fig:tetherShape}). The analytical estimated force and surface
tension are given as \cite{Derenyi2002}:
\begin{equation}
F = 2\pi\mathcal{K}_C/r,
\end{equation}  
and the surface tension
\begin{equation}
\alpha = 0.5\mathcal{K}_C/r^2,
\end{equation}  

\begin{table}
\centering
\begin{tabular}{l|c|c|c|c|c|c}          \hline
\multicolumn{7}{c}{Force vs. end-to-end distance} \\ \hline\hline
End-to-end distance ($\mu$m)              &6.8 & 8.2   & 9.2  & 10.2  & 11.6  & 12.8                \\  \hline
Computed tether radius ($\mu$m)           &n/a & n/a   & 0.20 & 0.165 & 0.140 & 0.105  \\  \hline
Computed force (pN)                       &0   & 1.41  & 1.76 & 2.14 & 2.72 & 3.68      \\  \hline
Analytical estimated force (pN)           &n/a & n/a   & 1.88 & 2.29 & 2.69 & 3.59      \\  \hline
Computed tension (pN/$\mu$m)              &0.05 & 0.43  & 0.71 & 0.98 & 1.60 & 2.93       \\ \hline 
Analytical estimated tension (pN/$\mu$m)  &n/a & n/a   & 0.75 & 1.10 & 1.53 & 2.72       
\end{tabular}
\caption{\label{table:forceEst} Computed and analytical estimated
  forces for each tethered shape. For the shape in
  Fig. \ref{fig:tether0.9}(c), the computed tether radius is not
  uniform along the extension direction, and $r=0.20 \mu$m is an
  estimate. As noted in the text a bending modulus of
  $\mathcal{K}_C=15K_bT$ is used for all calculations.}
 \end{table}

As Table \ref{table:forceEst} shows, for well developed tethered
shapes (end-to-end distance 11.6 and 12.8 $\mu$m, vesicle (e) and (f)
in Fig. \ref{fig:tether0.9}), the computed results and analytical
estimations are very close. It is a notable advantage that the present
simulation framework is also capable of force-extension calculations
for shapes that are not as simple as the schematic in
Fig.\ref{fig:tetherShape}.  Although the present example is in fact
axisymmetric, the algorithms are fully three-dimensional and can be
applied to loadings and shapes lacking symmetry.


\subsection{Lipid Phase Separation} \label{sec:phaseSeparation}
Membranes formed from different lipids can separate into distinct
domains (phases) according to their chemical properties, leading to
the formation of buds \cite{Baumgart2003, Bacia2005}. Baumgart et
al. \cite{Baumgart2005} found that their experiments are in good
agreement with line tension theory \cite{Lipowsky1992, Julicher1993,
  Julicher1996}, which treats domain interfaces as discrete with an
interface energy proportional to their length.  An alternative,
smooth-interface approach, based on traditional Ginzburg-Landau (GL)
theory \cite{BalluffiKoM, ToledanoLTPT} can be used to also model
phase separation \cite{Kawakatsu1993, Kawakatsu1994, Taniguchi1996,
  Ayton2005, Jiang2000}. One major drawback of the line tension model
is that it requires the system to be pre-phase-separated into
well-defined domains, preventing the consideration of composition
dynamics.

Here a GL model for a multi-component bilayer with two different
lipids in equilibrium is formulated, assuming that the vesicle is
composed of a mixture of two lipids denoted $A$ and $B$.  In general,
these two lipid types may have different constitutive properties, as
modeled by separate constitutive parameters: $\{\mathcal{K}_C^{(A)},
\mathcal{K}_G^{(A)}, C_0^{(A)} \}$ for lipid $A$, and
$\{\mathcal{K}_C^{(B)}, \mathcal{K}_G^{(B)},C_0^{(B)} \}$ for lipid
$B$.

Let the local concentrations of the two lipids be described by the
concentration parameters $c^{(A)}, c^{(B)} \in [0,1]$ with $c^{(A)} +
c^{(B)} = 1$.  The local lipid concentration at point $\mathsf{s} =
(s^1, s^2)$ can then be described by an order-parameter field
$c(\mathsf{s}) \equiv c^{(A)} $, which is referred to as the
concentration field or phase field.  The local constitutive properties
of the membrane can then be modeled as functions of the phase field
with convex combinations of the pure phase parameters:
\begin{subequations}\label{eq:variableModuli}
\begin{align}
\mathcal{K}_C &= 
c\mathcal{K}_C^{(A)} + (1-c)\mathcal{K}_C^{(B)}
\\
\mathcal{K}_G &= 
c\mathcal{K}_G^{(A)} + (1-c)\mathcal{K}_G^{(B)}
\\
C_0 &= 
c C_0^{(A)} + (1-c)C_0^{(B)} 
\end{align}
\end{subequations}
Thus rewriting the strain energy including explicit dependence of
fields on surface position,
\begin{equation}
E = \int_\mathcal{M}
\left\{
\frac{1}{2}\mathcal{K}_C(\mathsf{s}) [2 H(\mathsf{s}) - C_0(\mathsf{s})]^2 +
\mathcal{K}_G(\mathsf{s}) K(\mathsf{s})
\right\}
\sqrt{a} {d^2s}
\end{equation}
where explicit dependence of the mechanical properties on surface
coordinates $\mathsf{s}$ has been noted as a reminder of the
heterogeneity of the system.

The mechanics of the membrane are then dependent on both the shape of
the vesicle and the lipid composition.  Minimization of the total
potential energy now yields two sets of Euler-Lagrange equations, one
being the equilibrium equations related to variations in the shape
$\delta\bm{x}$, and the other being a phase equilibrium equation
related to variations in the concentration $\delta c$.

\begin{figure}
\centering
\includegraphics[height=1.5in]{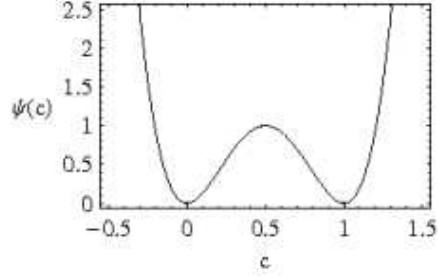}
\caption{\label{fig:doubleWell} Double-well normalized GL energy
  $\psi({c}) = 16c^2(c-1)^2$ used to model phase segregation in a
  two-component lipid system.}
\end{figure}

One further modification to the energy functional is needed to build
into the model of the physics of phases separation
\cite{ToledanoLTPT}.
\begin{equation} \label{eq:twoPhaseesEnergy}
I = E + \int_\mathcal{M} \Delta E [\psi({c}) + \epsilon^2|\nabla{c}|^2]\sqrt{a}d^2s .
\end{equation}
Here the normalized GL energy $\psi({c})$ is a double-well potential such as  
\[
\psi({c}) = 16c^2(c-1)^2
\] 
(see Fig. \ref{fig:doubleWell}) which is minimized when the
concentration $c$ takes a value of either 0 or 1, corresponding to
local lipid concentration of either pure type $A$ or pure type $B$.
The parameter $\Delta E$ scales the height of the barrier between the
two minima of $\psi({c})$, and controls the energy cost of a domain
interface.  The second addition to the energy describes short-range
cooperativity between neighboring lipids.  The parameter $\epsilon$ is
essentially a length scale which will determine the width of the
region of transition between phases.  As $\epsilon$ decreases to zero,
this region will limit to a curve where the concentration gradient can
be non-zero.  Inclusion of this penalty term in the energy will then
produce the effect of a diffuse line tension in the transition between
regions of pure phases.

\begin{figure}[tbp]
\centering
\subfigure[Simulation]{
  \includegraphics[width=0.45\textwidth]{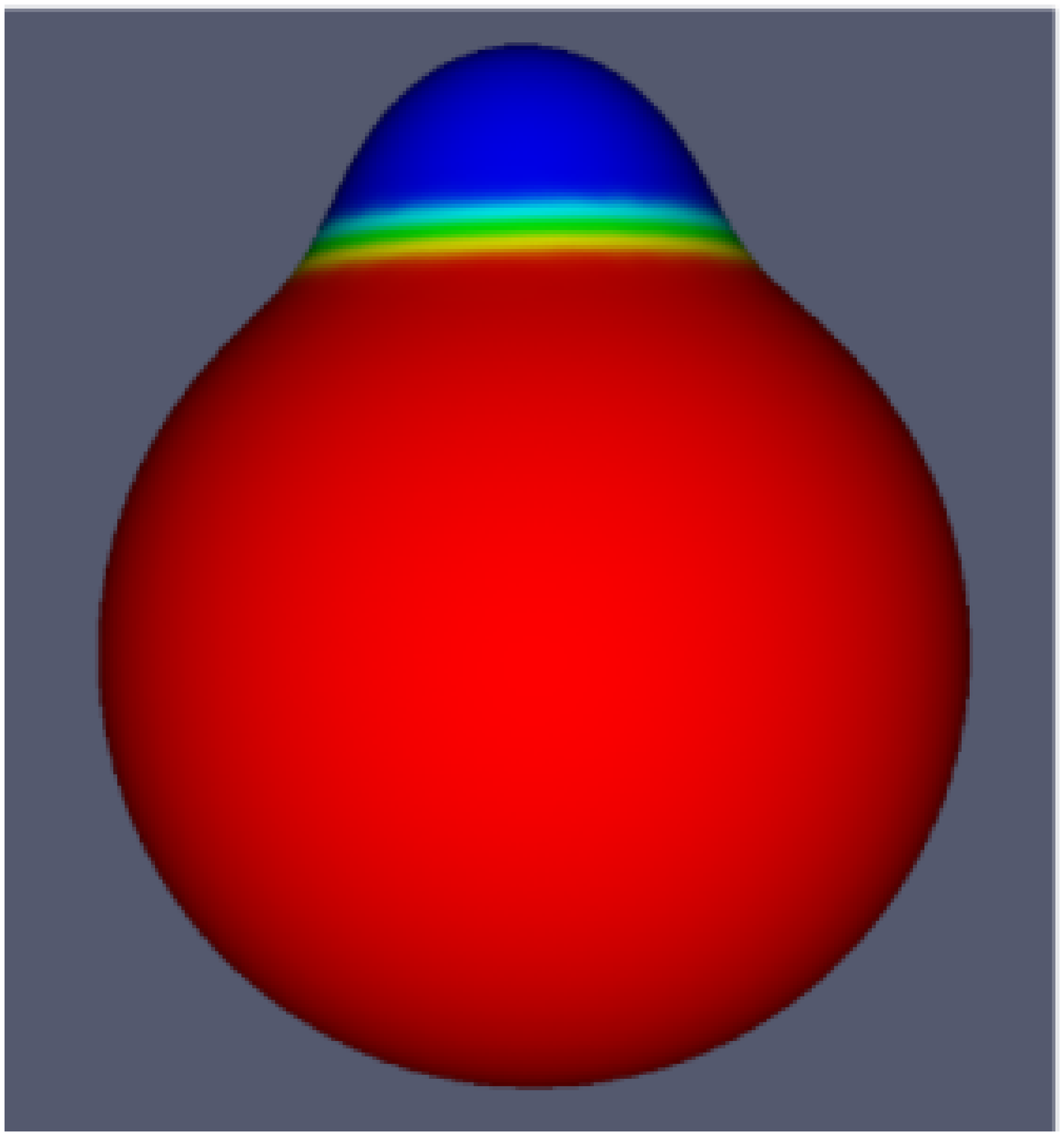}
} 
\subfigure[Experiment (Fig. 2G from \cite{Baumgart2005})]{
  \includegraphics[width=0.48\textwidth]{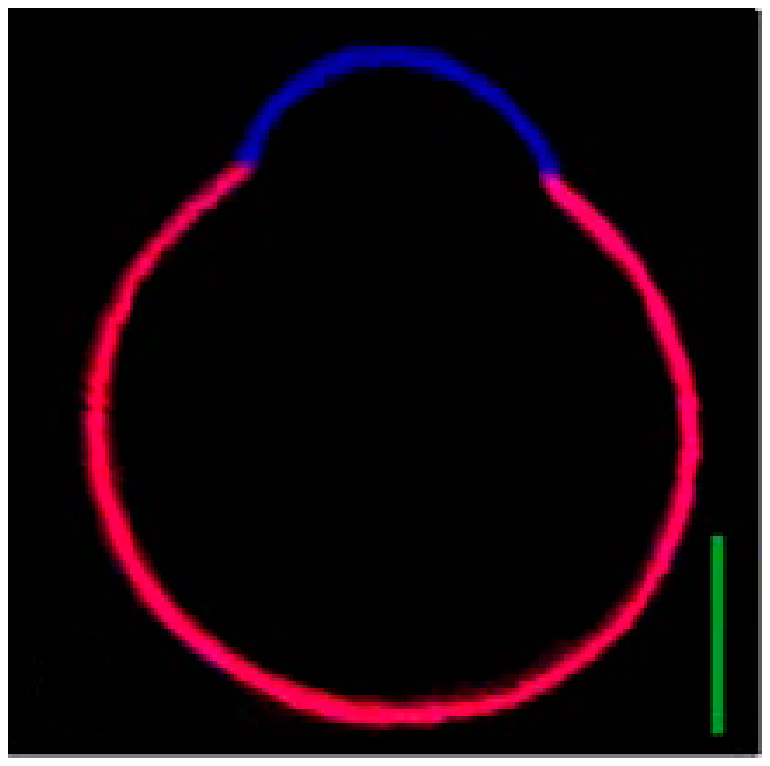}
} 
\subfigure[Simulation]{
  \includegraphics[width=0.45\textwidth]{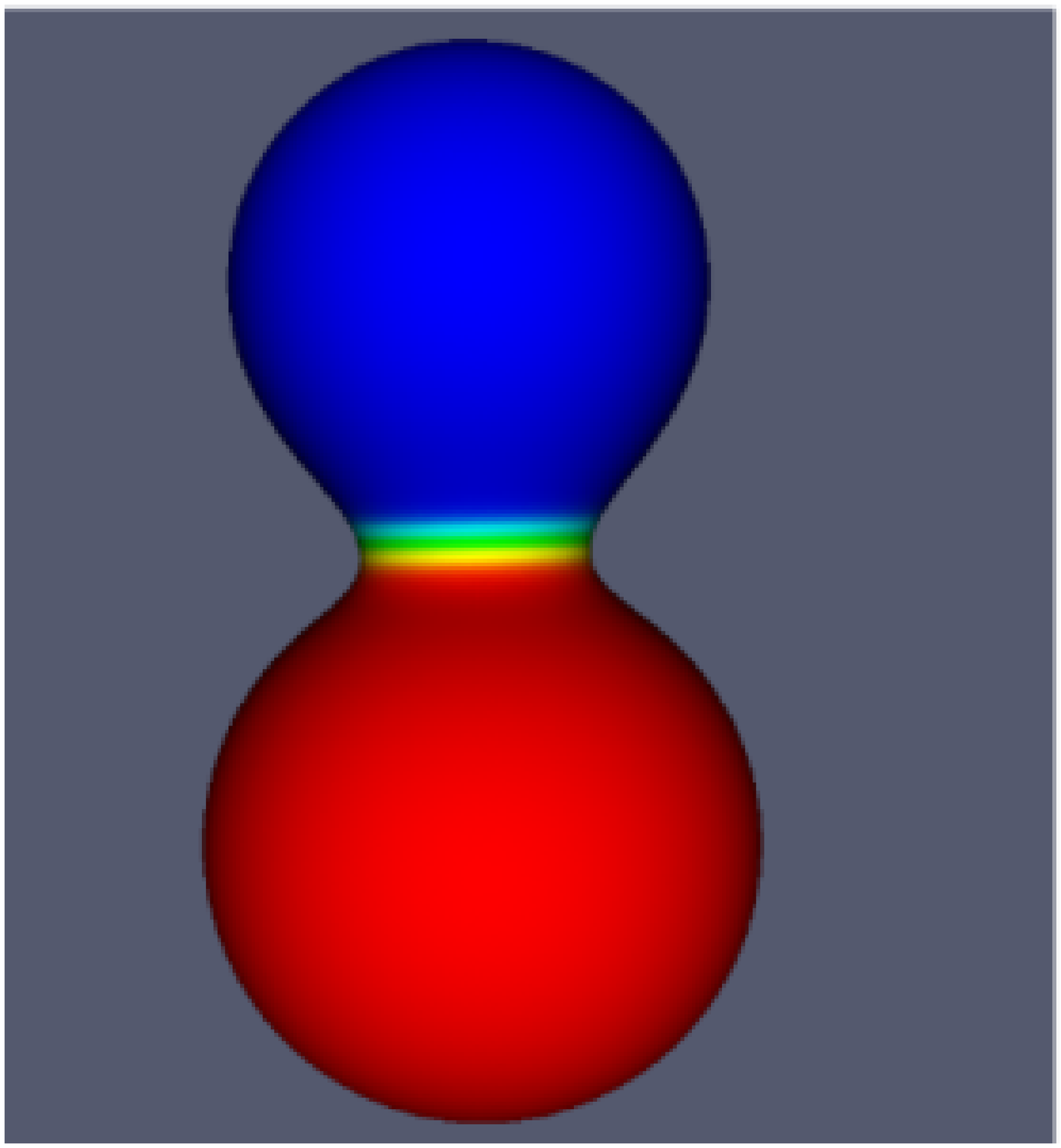}
} 
\subfigure[Experiment (Fig. 2A from \cite{Baumgart2005})]{
  \includegraphics[width=0.49\textwidth]{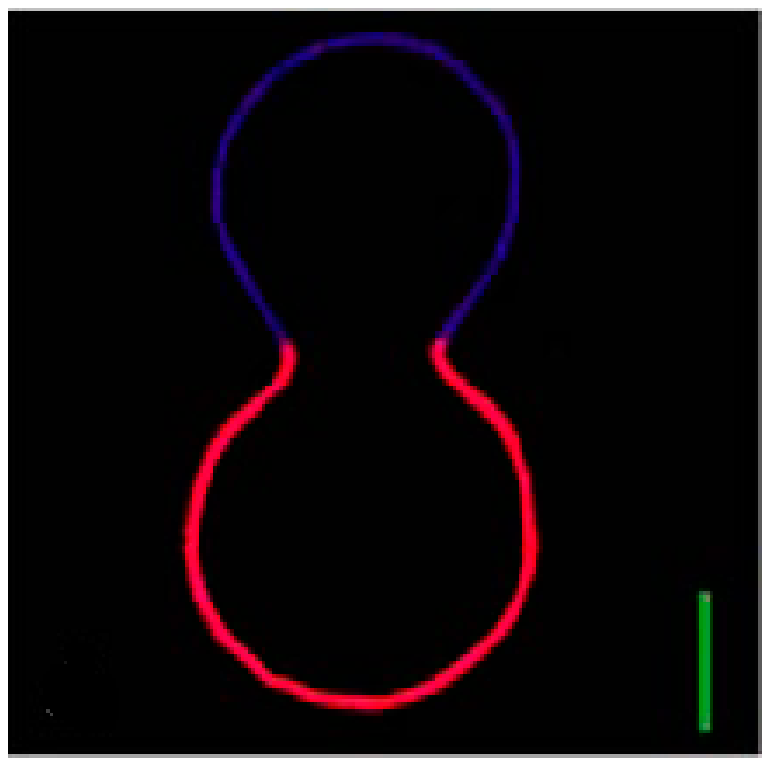}
} 
\caption{\label{fig:ps}Comparison of simulation and experiment for two
component lipid phase separation.  (a) \& (b) have reduced volume $\nu
= 0.98$ and global concentration $x_B=0.98$. (c) \& (d) have $\nu =
0.76$ and $x_B=0.56$.  Phase $A$ ($c=0$) is colored blue; phase $B$
($c=1$) is colored red.  Scale bars are $5 \mu$m.  Experimental images
are taken from the work of \citet{Baumgart2005}.}
\end{figure}

In Baumgart's experiment \cite{Baumgart2003, Baumgart2005}, bending
modulus $\mathcal{K}_C \approx 10^{-19}J$, line tension $\sigma
\approx 10^{-12} N$, and the radius of the vesicle $R \approx 10 \mu
m$.  Two vesicles from \cite{Baumgart2005} are simulated: one with
reduced volume $\nu = 0.98$, phase $B$ area fraction $x_{B}=(1/A)\int
c dA = 0.89$; the other $\nu = 0.76$, $x_{B}=(1/A)\int c dA = 0.56$,
starting from the original spherical shape with two separate domains
(Fig. 2A and 2G in \cite{Baumgart2005}). For the first one ($\nu =
0.98$), the simulation captured the small cap seen in the experiment
(Fig. \ref{fig:ps}).

\begin{figure}[ht]
\centering
\subfigure[standard dashpot regularization]{
  \includegraphics[height=3in]{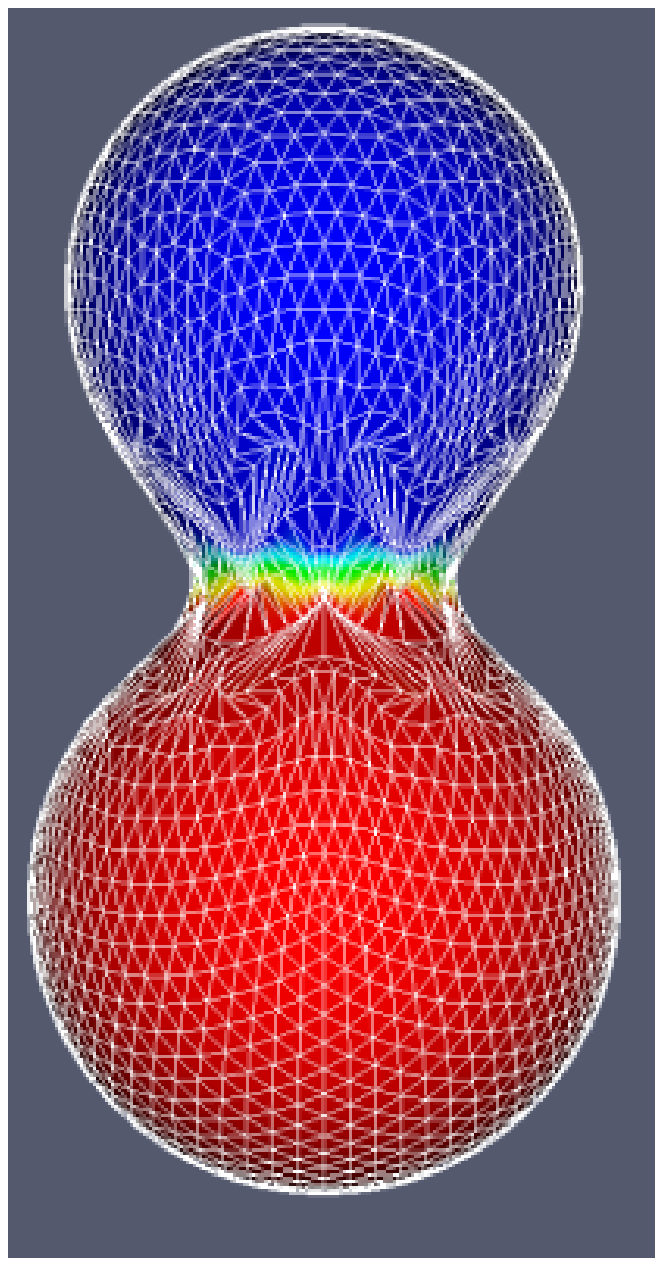}
}
\subfigure[r-adaptive regularization]{
  \includegraphics[height=3in]{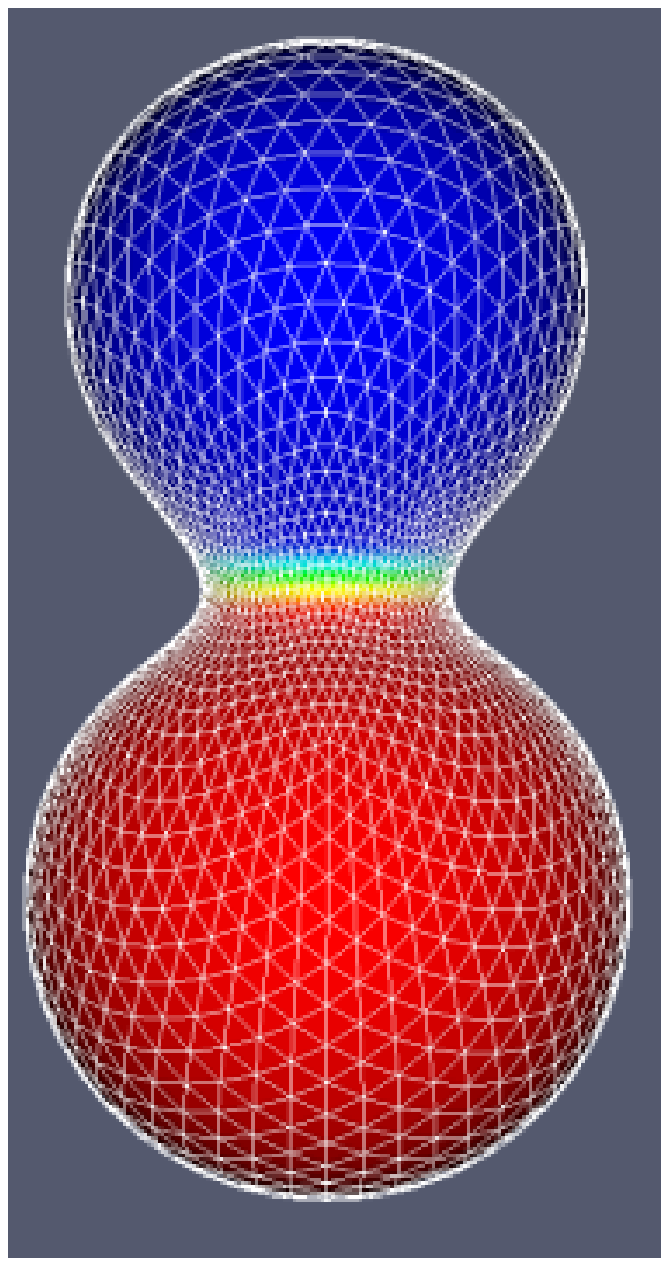}
}
\caption{\label{fig:nu.76mesh} r-adaptive regularization helps elements
perform well in the interface ($\nu =0.76$).}
\end{figure}

But for $\nu = 0.76$, the mesh around the interface is distorted
(Fig. \ref{fig:nu.76mesh} (a)). Because the simulation starts from a
sphere with roughly equilateral triangle elements, the shape change of
vesicle causes the elements in the interface region to contract
severely in the circumferential direction. This Element distortion
needs to be suppressed because it can lead to inaccuracy and
instability of the finite element simulation.

In order to get a good mesh after the deformation, the elements near
the interface need to contract in all directions so that they remain
equilateral, resulting in a greater \emph{density} of elements than
other parts of the vesicle. To tackle such large deformations,
remeshing strategies are often needed. The viscous regularization
introduced in section \ref{sec:stabilization} makes an r-adaptive
remeshing possible, since reference configuration can be arbitrarily
formulated to reposition the nodes of the mesh.  Here, a slight
modification of the dashpot regularization method is proposed with a
reference updating strategy that drives elements toward equilateral
shape.

Given an element of the mesh at regularization iteration
$n-1$ with area $A_{n-1}$,
r-adaptive regularization at step $n$ is defined by placing springs on
the three edges all of the same reference length
\[
\bar{\ell}_{n-1}= 2 \sqrt{\frac{A_{n-1}}{\sqrt{3}}} ,
\]
i.e., the length of a side of an equilateral triangle of the same area
$A_{n-1}$.
Thus the regularization energy term for each triangle is written as
\begin{equation}
E_n= \frac{k}{2}
\sum_{i=1}^3
(\ell_n^i-\bar{\ell}_{n-1})^2 ,
\end{equation} 
where the $\ell^i$ are the lengths of the element edges.

In principle this regularization energy could be applied to every
element in a mesh.  However, in practice these iterative updates are
slow to converge to a fully relaxed state (with zero regularization
energy), and depending on the regularization constant $k$ the method
can get stuck in a state with finite energy stored in the springs.
Hence, this ``equilateral'' form of the dashpot regularization is
only applied selectively to poorly shape elements, all the other
elements with the standard viscous regularization.
A shape-criteria $\gamma$ is then formulated to calculate different
spring energy for different elements,
\[
\gamma = \sum_{i=1}^3\frac{(\ell^i-\bar{\ell})^2}{\bar{\ell}^2}.
\]
Using this measure of shape quality, the regularization energy is
defined for each triangle by
\[
E_n = \begin{cases}
 \frac{k}{2}
\sum_{i=1}^3
(\ell_n^i-\bar{\ell}_{n-1})^2, \quad \gamma \text{ large},\\
 \frac{k}{2}
\sum_{i=1}^3
(\ell_n^i-\ell_{n-1}^i)^2, \quad \gamma \text{ small} .
\end{cases}
\]
In other words, if $\gamma$ is large (say, $\gamma>1$) for an element,
it has poor shape and r-adaptive regularization is used on that
element; if $\gamma$ is small enough, reference lengths are updated
from the deformed lengths of the previous iteration as for the dashpot
model described earlier.  The addition of r-adaptive regularization
has the effect of moving the nodes around on the membrane surface.  In
the present example of a phase-separated vesicle, this results in a
finer mesh near the interface area than elsewhere
(Fig. \ref{fig:nu.76mesh} (b)). For reduced volume $\nu=0.76$, the
simulated result is shown in Fig. \ref{fig:ps} to compare with the
experimental result.

\section{Conclusions}\label{sec:conclusions}
In this paper a framework for three-dimensional analysis of mechanics
of lipid bilayer membranes is presented, based on the finite element
method. Particular interest is focused on large deformation problems:
tether formation (Sec. \ref{sec:tetherFormation}) and phase separation
(Sec. \ref{sec:phaseSeparation}).

The primary difficulty faced in FE simulation of fluid membranes is
the presense of mesh instabilities linked to the
parameterization-independent nature of fluid surfaces. Curvature
models of vesicle mechanics depend only on current shape, and thus is
not sensitive to in-plane (stretching and shearing) deformations of
the surface FE mesh. Here a viscous regularization method is thus
introduced to regularize tangential mesh deformations. In this method
artificial reference configurations and corresponding in-plane
energies are added to stabilize the tangential deformations; reference
updates are designed so that artificial energy converges to zero in
order to retain the physics of the original model.

Regulariztion of tangential mesh deformations eliminates the need for
local enforcement of membrane incompressibility \citep{Feng2006},
providing a more convenient setting for augmented Lagrangian (AL)
enforcement of global constraints on area and volume. The AL method
can achieve higher accuracy with lower computational cost, compared to
the penalty method. 

Large deformation problems can be very sensitive to mesh quality. 
Because of the physical meaninglessness of the reference
configurations in the simulation, r-adaptive remeshing is easy to
achieve in the context of viscous regularization, simply choosing a
reference updating strategy which will reposition the nodes to get a
better quality mesh.

One promising direction for future work is to combine viscous
regularization and r-adaptive remeshing with the dynamic triangulation
approach \cite{GompperKroll1997}, in which the edge of a pair of
triangles swaps to form less distorted triangle elements instantly.
This could be a powerful approach, speeding up the otherwise slow
movements of nodes driven by the viscous regularization. Also, the
success of r-adaptive regularization relies to some degree on the
quality of the starting mesh. If there are too many badly shaped
element and the shape criteria $\gamma$ tolerance is chosen to be too
small, the mesh can sometimes lock with non-zero regularization
energy, resulting in physically wrong shapes of vesicles. Dynamic
triangulation could aleviate such locking.  Lastly a r-adaptive
regularized dynamic triangulation strategy could avoid the need for
global remeshing in large deformation problems such as the tether
simulations in Sec. \ref{sec:tetherFormation}.


Although the problems simulated in this paper are all axisymmetric,
the model is really designed for fully three-dimensional calculations, and can
thus deal with arbitrary geometries and loads. 
For example future applications such as mechanics of organelles like
mitochondria \cite{Frey2000} and endoplasmic reticulum (ER)
\cite{Snapp2003}, with incredibly complex shapes may provide exciting
opportunities for future study with these methods.

\bibliographystyle{unsrtnat}
\bibliography{references}

\end{document}